\documentclass{book}

\usepackage[linktocpage]{hyperref}
\usepackage{fullpage}
\usepackage{makeidx}
\usepackage{graphicx}

\newcommand{\xs}[1]{Section~\ref{#1}}

\newcommand{\C}{{C\index{C}}}
\newcommand{\cpp}{{C++\index{C++}}}
\newcommand{\csharp}{{C\#\index{C\#}}}
\newcommand{\perl}{{Perl\index{Perl}}}
\newcommand{\java}{{Java\index{Java}}}
\newcommand{\python}{{Python\index{Python}}}
\newcommand{\fortran}{{Fortran\index{Fortran}}}
\newcommand{\aspectj}{{AspectJ\index{AspectJ}}}
\newcommand{\php}{{PHP\index{PHP}}}

\newcommand{\lisp}{{LISP\index{LISP}}}
\newcommand{\scheme}{{Scheme\index{Scheme}}}
\newcommand{\haskell}{{Haskell\index{Haskell}}}
\newcommand{\mllessequal}{{ML$_{\le}$\index{ML$_{\le}$}}}
\newcommand{\fcpp}{{FC++\index{FC++}}}

\newcommand{\file}[1]{\url{#1}\index{Files!#1}}
\newcommand{\tool}[1]{\texttt{#1}\index{Tools!#1}}

\newcommand{\api}[1]{\texttt{#1}\index{API!#1}}

\newcommand{\javacc}[0]{JavaCC\index{Tools!JavaCC}}

\newcommand{\macos}[1]{\index{Mac OS #1@{\sc{Mac OS #1}}}{\sc{Mac OS #1}}}

\newcommand{\lucidL}[1]{{$\mathit{Lucid}$}($L$) }

		{}

\def\myvert{\raise 2.27pt \hbox{\vrule depth 0pt height 8pt width 0.2mm}}
\def\myarrow{\hspace*{0.43mm}%
             \raise 2.29pt\hbox{\vrule depth 0pt height 8pt width 0.16mm}%
             \hspace*{-0.32mm}%
             $\longrightarrow$
             \ %
             }

\makeindex

\begin{document}

\title{Comparative Studies of Programming Languages, COMP6411 Lecture Notes, \\Revision 1.9}
\author{Joey Paquet \and Serguei A. Mokhov (Eds.)}

\maketitle

\chapter*{Preface}

{\em Lecture notes for the Comparative Studies of Programming Languages course, COMP6411,
taught at the Department of Computer Science and Software Engineering, Faculty
of Engineering and Computer Science, Concordia University, Montreal, QC, Canada.
These notes include a compiled book of primarily related articles from the
Wikipedia, the Free Encyclopedia \cite{wikipedia}, as well as Comparative Programming Languages book \cite{comparative-pls-3rd}
and other resources, including our own. The original notes were compiled by Dr. Paquet \cite{paquet-comp6411-w10-lecture-notes}}

\tableofcontents

\chapter{Brief History and Genealogy of Programming Languages}

\section{Introduction}

A programming language is an artificial language designed to express computations that can be performed by a machine, particularly a computer. Programming languages can be used to create programs that control the behavior of a machine, to express algorithms precisely, or as a mode of human communication.
The earliest programming languages predate the invention of the computer, and were used to direct the behavior of mechanical machines such as player pianos. Thousands of different programming languages have been created, mainly in the computer field, with many more being created every year. Most programming languages describe computation in an imperative style, i.e., as a sequence of commands, although some languages, such as those that support functional programming or logic programming, use alternative forms of description.

\subsection{Subreferences}
\begin{enumerate}
	\item 
\url{http://en.wikipedia.org/wiki/Programming_language}
\end{enumerate}

\section{History}
\index{History}

\subsection{Pre-computer era}
\index{History!Pre-computer era}

\subsubsection{Analytical engine}
\index{History!Analytical engine}

The analytical engine, an important step in the history of computers, was the design of a mechanical general-purpose computer by the British mathematician Charles Babbage. It was first described in 1837. Because of financial, political, and legal issues, the engine was never built. In its logical design the machine was essentially modern, anticipating the first completed general-purpose computers by about 100 years.
The input (programs and data) was to be provided to the machine via punched cards, a method being used at the time to direct mechanical looms such as the Jacquard loom. For output, the machine would have a printer, a curve plotter and a bell. The machine would also be able to punch numbers onto cards to be read in later. It employed ordinary base-10 fixed-point arithmetic.
There was to be a store (i.e., a memory) capable of holding 1,000 numbers of 50 decimal digits each (ca. 20.7kB). An arithmetical unit (the ``mill") would be able to perform all four arithmetic operations, plus comparisons and optionally square roots. Like the central processing unit (CPU) in a modern computer, the mill would rely upon its own internal procedures, to be stored in the form of pegs inserted into rotating drums called ``barrels,'' in order to carry out some of the more complex instructions the user's program might specify.
The programming language to be employed by users was akin to modern day assembly languages. Loops and conditional branching were possible and so the language as conceived would have been Turing-complete long before Alan Turing's concept. Three different types of punch cards were used: one for arithmetical operations, one for numerical constants, and one for load and store operations, transferring numbers from the store to the arithmetical unit or back. There were three separate readers for the three types of cards.

\subsection{Subreferences}

\begin{enumerate}
	\item 
\url{http://en.wikipedia.org/wiki/Analytical_engine}
\end{enumerate}

\subsection{Early computer era}
\index{History!Early computer era}

\subsubsection{Plankalkül}
\index{Plankalkül}

Plankalkül is a computer language developed for engineering purposes by Konrad Zuse. It was the first high-level non-von Neumann programming language to be designed for a computer and was designed between 1943 and 1945. Also, notes survive with scribblings about such a plan calculation dating back to 1941. Plankalkül was not published at that time owing to a combination of factors such as conditions in wartime and postwar Nazi Germany. By 1946, Zuse had written a book on the subject but this remained unpublished. In 1948 Zuse published a paper about the Plankalkül in the ``Archiv der Mathematik'' but still did not attract much feedback - for a long time to come programming a computer would only be thought of as programming with machine code. The Plankalkül was eventually more comprehensively published in 1972 and the first compiler for it was implemented in 1998. Another independent implementation followed in the year 2000 by the Free University of Berlin.
Plankalkül drew comparisons to APL and relational algebra. It includes assignment statements, subroutines, conditional statements, iteration, floating point arithmetic, arrays, hierarchical record structures, assertions, exception handling, and other advanced features such as goal-directed execution. Thus, this language included many of the syntactical elements of structured programming languages that would be invented later, but it failed to be recognized widely. 

\subsection{Subreferences}

\begin{enumerate}
	\item 
\url{http://en.wikipedia.org/wiki/Plankalk%C3%BCl}
\end{enumerate}

\subsubsection{Short Code}
\index{Short Code}

Short Code was one of the first higher-level languages ever developed for an electronic computer. Unlike machine code, Short Code statements represented mathematic expressions rather than a machine instruction.
Short Code was proposed by John Mauchly in 1949 and originally known as Brief Code. William Schmitt implemented a version of Brief Code in 1949 for the BINAC computer, though it was never debugged and tested. The following year Schmitt implemented a new version of Brief Code for the UNIVAC I where it was now known as Short Code. 
While Short Code represented expressions, the representation itself was not direct and required a process of manual conversion. Elements of an expression were represented by two-character codes and then divided into 6-code groups in order to conform to the 12 byte words used by BINAC and Univac computers.
Along with basic arithmetic, Short Code allowed for branching and calls to a library of functions. The language was interpreted and ran about 50 times slower than machine code.

\begin{enumerate}
	\item 
\url{http://en.wikipedia.org/wiki/Short_Code_%28computer_language%29}
\end{enumerate}

\subsubsection{A-0}

The A-0 system (Arithmetic Language version 0), written by Grace Hopper in 1951 and 1952 for the UNIVAC I, was the first compiler ever developed for an electronic computer. The A-0 functioned more as a loader or linker than the modern notion of a compiler. A program was specified as a sequence of subroutines and arguments. The subroutines were identified by a numeric code and the arguments to the subroutines were written directly after each subroutine code. The A-0 system converted the specification into machine code that could be fed into the computer a second time to execute the program. The A-0 system was followed by the A-1, A-2, A-3 (released as ARITH-MATIC), AT-3 (released as MATH-MATIC) and B-0 (released as FLOW-MATIC).

\begin{enumerate}
	\item 
\url{http://en.wikipedia.org/wiki/A-0_System}
\end{enumerate}

\subsection{Modern/Structured programming languages}
\index{History!Modern/Structured programming languages}

\subsubsection{Fortran}

Fortran (previously FORTRAN) is a general-purpose, procedural, imperative programming language that is especially suited to numeric computation and scientific computing. Originally developed by IBM in the 1950s for scientific and engineering applications, Fortran came to dominate this area of programming early on and has been in continual use for over half a century in computationally intensive areas such as numerical weather prediction, finite element analysis, computational fluid dynamics (CFD), computational physics, and computational chemistry. It is one of the most popular languages in the area of high-performance computing and is the language used for programs that benchmark and rank the world's fastest supercomputers.

Fortran (a blend derived from The IBM Mathematical Formula Translating System) encompasses a lineage of versions, each of which evolved to add extensions to the language while usually retaining compatibility with previous versions. Successive versions have added support for processing of character-based data (FORTRAN 77), array programming, modular programming and object-based programming (Fortran 90 / 95), and object-oriented and generic programming (Fortran 2003).

\begin{enumerate}
	\item 
\url{http://en.wikipedia.org/wiki/Fortran}
\end{enumerate}

\subsubsection{Algol}

ALGOL (short for ALGOrithmic Language) is a family of imperative computer programming languages originally developed in the mid 1950s which greatly influenced many other languages and became the de facto way algorithms were described in textbooks and academic works for almost the next 30 years. It was designed to avoid some of the perceived problems with FORTRAN and eventually gave rise to many other programming languages (including BCPL, B, Pascal, Simula and C). ALGOL introduced code blocks and was the first language to use begin and end pairs for delimiting them.

John Backus developed the Backus normal form method of describing programming languages specifically for ALGOL 58. It was revised and expanded by Peter Naur for ALGOL 60, and at Donald Knuth's suggestion renamed Backus–Naur Form.
Niklaus Wirth based his own ALGOL W on ALGOL 60 before moving to develop Pascal. 

\begin{enumerate}
	\item 
\url{http://en.wikipedia.org/wiki/ALGOL}
\end{enumerate}

\subsubsection{{\lisp}}

{\lisp} is a family of computer programming languages with a long history and a distinctive, fully parenthesized syntax. Originally specified in 1958, Lisp is the second-oldest high-level programming language in widespread use today; only Fortran is older. Like Fortran, Lisp has changed a great deal since its early days, and a number of dialects have existed over its history. Today, the most widely known general-purpose Lisp dialects are Clojure, Common Lisp and {\scheme}.
Lisp was originally created as a practical mathematical notation for computer programs, influenced by the notation of Alonzo Church's lambda calculus. It quickly became the favored programming language for artificial intelligence (AI) research. As one of the earliest programming languages, Lisp pioneered many ideas in computer science, including tree data structures, automatic storage management, dynamic typing, and the self-hosting compiler.
The name LISP derives from ``LISt Processing''. Linked lists are one of Lisp languages' major data structures, and Lisp source code is itself made up of lists. As a result, Lisp programs can manipulate source code as a data structure, giving rise to the macro systems that allow programmers to create new syntax or even new domain-specific programming languages embedded in Lisp.

\begin{enumerate}
	\item 
\url{http://en.wikipedia.org/wiki/Lisp_%28programming_language%29}
\end{enumerate}

\subsubsection{Cobol}

COBOL is one of the oldest programming languages. Its name is an acronym for COmmon Business-Oriented Language, defining its primary domain in business, finance, and administrative systems for companies and governments.
A specification of COBOL was initially created during the second half of 1959 by Grace Hopper. The specifications were to a great extent inspired by the FLOW-MATIC language invented by Grace Hopper, commonly referred to as ``the mother of the COBOL language''. 
Since 1959 COBOL has undergone several modifications and improvements. In an attempt to overcome the problem of incompatibility between different versions of COBOL, the American National Standards Institute (ANSI) developed a standard form of the language in 1968. This version was known as American National Standard (ANS) COBOL. In 1974, ANSI published a revised version of (ANS) COBOL, containing a number of features that were not in the 1968 version. In 1985, ANSI published still another revised version that had new features not in the 1974 standard. The language continues to evolve today. The COBOL 2002 standard includes support for object-oriented programming and other modern language features.

\begin{enumerate}
	\item 
\url{http://en.wikipedia.org/wiki/COBOL}
\end{enumerate}

\subsubsection{Simula}

Simula is a name for two programming languages, Simula I and Simula 67, developed in the 1960s at the Norwegian Computing Center in Oslo, by Ole-Johan Dahl and Kristen Nygaard. Syntactically, it is a fairly faithful superset of Algol 60. 
Simula 67 introduced objects, classes, subclasses, virtual methods, coroutines, discrete event simulation, and features garbage collection.

Simula is considered the first object-oriented programming language. As its name implies, Simula was designed for doing simulations, and the needs of that domain provided the framework for many of the features of object-oriented languages today.
Simula has been used in a wide range of applications such as simulating VLSI designs, process modeling, protocols, algorithms, and other applications such as typesetting, computer graphics, and education. Since Simula-type objects are reimplemented in {\cpp}, Java and {\csharp} the influence of Simula is often understated. The creator of {\cpp}, Bjarne Stroustrup, has acknowledged that Simula 67 was the greatest influence on him to develop {\cpp}, to bring the kind of productivity enhancements offered by Simula to the raw computational speed offered by lower level languages like BCPL.

\begin{enumerate}
	\item 
\url{http://en.wikipedia.org/wiki/Simula}
\end{enumerate}

\subsubsection{Basic}

In computer programming, BASIC (an acronym for Beginner's All-purpose Symbolic Instruction Code) is a family of high-level programming languages. The original BASIC was designed in 1964 by John George Kemeny and Thomas Eugene Kurtz at Dartmouth College in New Hampshire, USA to provide computer access to non-science students. At the time, nearly all use of computers required writing custom software, which was something only scientists and mathematicians tended to be able to do. The language and its variants became widespread on microcomputers in the late 1970s and 1980s. The language was based partly on FORTRAN II and partly on ALGOL 60. 

It was intended to address the complexity issues of older languages with a new language design specifically for the new class of users that the then new time-sharing systems allowed—that is, a less technical user who did not have the mathematical background of the more traditional users and was not interested in acquiring it. Being able to use a computer to support teaching and research was quite novel at the time. 

The eight original design principles of BASIC were:

\begin{itemize}
	\item 
Be easy for beginners to use.
	\item 
Be a general-purpose programming language.
	\item 
Allow advanced features to be added for experts (while keeping the language simple for beginners).
	\item 
Be interactive.
	\item 
Provide clear and friendly error messages.
	\item 
Respond quickly for small programs.
	\item 
Not to require an understanding of computer hardware.
	\item 
Shield the user from the operating system.
\end{itemize}

BASIC\index{BASIC} remains popular to this day in a handful of highly modified dialects and new languages influenced by BASIC such as Microsoft Visual Basic. As of 2006, 59\% of developers for the .NET platform used Visual Basic .NET as their only language.

\begin{enumerate}
	\item 
\url{http://en.wikipedia.org/wiki/BASIC}
\end{enumerate}

\subsubsection{ISWIM}
\index{ISWIM}

ISWIM \cite{ISWIM} is an abstract computer programming language
(or a family of programming languages) devised by Peter J. Landin
and first described in his article, The Next 700 Programming Languages,
published in the Communications of the ACM in 1966 \cite{ISWIM}.
The acronym stands for ``If you See What I Mean''.
Although not implemented, it has proved very influential in the development
of programming languages, especially functional programming languages such as
SASL, Miranda, ML, Haskell and their successors.
A notationally distinctive feature of ISWIM is its use of \api{where} clauses.
An ISWIM program is a single expression qualified by \api{where} clauses
(auxiliary definitions including equations among variables),
conditional expressions and function definitions. With CPL, ISWIM was one
of the first programming languages to use \api{where} clauses,
which are used widely in functional and declarative programming languages.

\begin{enumerate}
	\item 
\url{http://en.wikipedia.org/wiki/ISWIM}
\end{enumerate}

\subsubsection{Pascal}
\index{Pascal}

Pascal is an influential imperative and procedural programming language, designed in 1968/9 and published in 1970 by Niklaus Wirth as a small and efficient language intended to encourage good programming practices using structured programming and data structuring.

Pascal is based on the ALGOL programming language and named in honor of the French mathematician and philosopher Blaise Pascal. Wirth subsequently developed the Modula-2 and Oberon, languages similar to Pascal. Before, and leading up to Pascal, Wirth developed the language Euler, followed by Algol-W.
Initially, Pascal was largely, but not exclusively, intended to teach students structured programming. Generations of students have used Pascal as an introductory language in undergraduate courses. A derivative known as Object Pascal was designed for object oriented programming.

\begin{enumerate}
	\item 
\url{http://en.wikipedia.org/wiki/Pascal_%28programming_language%29}
\end{enumerate}

\subsubsection{Smalltalk}
\index{Smalltalk}

Smalltalk is an object-oriented, dynamically typed, reflective programming language. Smalltalk was created as the language to underpin the ``new world'' of computing exemplified by ``human–computer symbiosis.'' It was designed and created in part for educational use at the Learning Research Group of Xerox PARC by Alan Kay, Dan Ingalls, Adele Goldberg, Ted Kaehler, Scott Wallace, and others during the 1970s.

The language was first generally released as Smalltalk-80 and has been widely used since. Smalltalk-like languages are in continuing active development, and have gathered loyal communities of users around them. ANSI Smalltalk was ratified in 1998 and represents the standard version of Smalltalk.

One of the early versions, Smalltalk-76, was one of the first programming languages to be implemented along with a development environment featuring most of the now familiar tools, including a class library code browser/editor. Smalltalk-80 added metaclasses, to help maintain the ``everything is an object'' (except private instance variables) paradigm by associating properties and behavior with individual classes. Smalltalk programs are usually compiled to bytecode, which is then interpreted by a virtual machine or dynamically translated into machine-native code as in just-in-time compilation.

Smalltalk has influenced the wider world of computer programming in four main areas. It inspired the syntax and semantics of other computer programming languages. Secondly, it was a prototype for a model of computation known as message passing. Thirdly, its WIMP GUI inspired the windowing environments of personal computers in the late twentieth and early twenty-first centuries, so much so that the windows of the first Macintosh desktop look almost identical to the MVC windows of Smalltalk-80. Finally, the integrated development environment was the model for a generation of visual programming tools that look like Smalltalk's code browsers and debuggers.

As in other object-oriented languages, the central concept in Smalltalk-80 (but not in Smalltalk-72) is that of an object. An object is always an instance of a class. Classes are ``blueprints'' that describe the properties and behavior of their instances. For example, a Window class would declare that windows have properties such as the label, the position and whether the window is visible or not. The class would also declare that instances support operations such as opening, closing, moving and hiding. Each particular Window object would have its own values of those properties, and each of them would be able to perform operations defined by its class.

A Smalltalk object can do exactly three things:

\begin{enumerate}
	\item 
  Hold state (references to other objects).
	\item 
  Receive a message from itself or another object.
	\item 
  In the course of processing a message, send messages to itself or another object.
\end{enumerate}

The state an object holds is always private to that object. Other objects can query or change that state only by sending requests (messages) to the object to do so. Any message can be sent to any object: when a message is received, the receiver determines whether that message is appropriate. (Alan Kay has commented that despite the attention given to objects, messaging is the most important concept in Smalltalk.)

Smalltalk is a ``pure'' object-oriented programming language, meaning that, unlike Java and {\cpp}, there is no difference between values which are objects and values which are primitive types. In Smalltalk, primitive values such as integers, booleans and characters are also objects, in the sense that they are instances of corresponding classes, and operations on them are invoked by sending messages. A programmer can change the classes that implement primitive values, so that new behavior can be defined for their instances--for example, to implement new control structures--or even so that their existing behavior will be changed. This fact is summarized in the commonly heard phrase ``In Smalltalk everything is an object''. Since all values are objects, classes themselves are also objects. Each class is an instance of the metaclass of that class. Metaclasses in turn are also objects, and are all instances of a class called Metaclass. Code blocks are also objects.

Smalltalk-80 is a totally reflective system, implemented in Smalltalk-80 itself. Smalltalk-80 provides both structural and computational reflection. Smalltalk is a structurally reflective system whose structure is defined by Smalltalk-80 objects. The classes and methods that define the system are themselves objects and fully part of the system that they help define. The Smalltalk compiler compiles textual source code into method objects, typically instances of CompiledMethod. These get added to classes by storing them in a class's method dictionary. The part of the class hierarchy that defines classes can add new classes to the system. The system is extended by running Smalltalk-80 code that creates or defines classes and methods. In this way a Smalltalk-80 system is a ``living'' system, carrying around the ability to extend itself at run time.

Since the classes are themselves objects, they can be asked questions such as ``what methods do you implement?'' or ``what fields/slots/instance variables do you define?''. So objects can easily be inspected, copied, (de)serialized and so on with generic code that applies to any object in the system.

Smalltalk-80 also provides computational reflection, the ability to observe the computational state of the system. In languages derived from the original Smalltalk-80 the current activation of a method is accessible as an object named via a keyword, thisContext. By sending messages to thisContext a method activation can ask questions like ``who sent this message to me''. These facilities make it possible to implement co-routines or Prolog-like back-tracking without modifying the virtual machine. 

When an object is sent a message that it does not implement, the virtual machine sends the object the doesNotUnderstand: message with a reification of the message as an argument. The message (another object, an instance of Message) contains the selector of the message and an Array of its arguments. In an interactive Smalltalk system the default implementation of doesNotUnderstand: is one that opens an error window reporting the error to the user. Through this and the reflective facilities the user can examine the context in which the error occurred, redefine the offending code, and continue, all within the system, using Smalltalk-80's reflective facilities.

\begin{enumerate}
	\item 
\url{http://en.wikipedia.org/wiki/Smalltalk}
\end{enumerate}

\subsubsection{C}

{\C} is a general-purpose computer programming language developed in 1972 by Dennis Ritchie at the Bell Telephone Laboratories for use with the Unix operating system.

Although C was designed for implementing system software, it is also widely used for developing portable application software.
C is one of the most popular programming languages. It is widely used on many different software platforms, and there are few computer architectures for which a C compiler does not exist. C has greatly influenced many other popular programming languages, most notably {\cpp}, which originally began as an extension to C.

C is an imperative (procedural) systems implementation language. It was designed to be compiled using a relatively straightforward compiler, to provide low-level access to memory, to provide language constructs that map efficiently to machine instructions, and to require minimal run-time support. C was therefore useful for many applications that had formerly been coded in assembly language. Despite its low-level capabilities, the language was designed to encourage machine-independent programming. A standards-compliant and portably written C program can be compiled for a very wide variety of computer platforms and operating systems with little or no change to its source code. The language has become available on a very wide range of platforms, from embedded microcontrollers to supercomputers.

C also exhibits the following more specific characteristics:

\begin{itemize}
	\item 
lack of nested function definitions
	\item 
variables may be hidden in nested blocks
	\item 
partially weak typing; for instance, characters can be used as integers
	\item 
low-level access to computer memory by converting machine addresses to typed pointers
	\item 
function and data pointers supporting ad hoc run-time polymorphism
	\item 
array indexing as a secondary notion, defined in terms of pointer arithmetic
	\item 
a preprocessor for macro definition, source code file inclusion, and conditional compilation
	\item 
complex functionality such as I/O, string manipulation, and mathematical functions consistently delegated to library routines
	\item 
A relatively small set of reserved keywords
\end{itemize}

C does not have some features that are available in some other programming languages:

\begin{itemize}
	\item 
No direct assignment of arrays or strings (copying can be done via standard functions; assignment of objects having struct or union type is supported)
	\item 
No automatic garbage collection
	\item 
No requirement for bounds checking of arrays
	\item 
No operations on whole arrays
	\item 
No syntax for ranges, such as the A..B notation used in several languages
	\item 
Prior to C99, no separate Boolean type (zero/nonzero is used instead)[6]
	\item 
No functions as parameters (only function and variable pointers)
	\item 
No exception handling; standard library functions signify error conditions with the global errno variable and/or special return values
	\item 
Only rudimentary support for modular programming
	\item 
No compile-time polymorphism in the form of function or operator overloading
	\item 
Only rudimentary support for generic programming
	\item 
Very limited support for object-oriented programming with regard to polymorphism and inheritance
	\item 
Limited support for encapsulation
	\item 
No native support for multithreading and networking
	\item 
No standard libraries for computer graphics and several other application programming needs
\end{itemize}

\begin{enumerate}
	\item 
\url{http://en.wikipedia.org/wiki/C_%28programming_language%29}
\end{enumerate}

\subsubsection{Prolog}
\index{Prolog}

Prolog is a general purpose logic programming language associated with artificial intelligence and computational linguistics.
Prolog has its roots in formal logic, and unlike many other programming languages, Prolog is declarative: The program logic is expressed in terms of relations, represented as facts and rules. Execution is triggered by running queries over these relations.
The language was first conceived by a group around Alain Colmerauer in Marseille, France, in the early 1970s and the first Prolog system was developed in 1972 by Alain Colmerauer and Phillipe Roussel.
Prolog was one of the first logic programming languages, and remains among the most popular such languages today, with many free and commercial implementations available. While initially aimed at natural language processing, the language has since then stretched far into other areas like theorem proving, expert systems, games, automated answering systems, ontologies and sophisticated control systems. Modern Prolog environments support the creation of graphical user interfaces, as well as administrative and networked applications.

\begin{enumerate}
	\item 
\url{http://en.wikipedia.org/wiki/Prolog}
\end{enumerate}

\subsubsection{ML}

ML is a general-purpose functional programming language developed by Robin Milner and others in the late 1970s at the University of Edinburgh, whose syntax is inspired by ISWIM. Historically, ML stands for metalanguage: it was conceived to develop proof tactics in the LCF theorem prover (whose language, pplambda, a combination of the first-order predicate calculus and the simply typed polymorphic lambda-calculus, had ML as its metalanguage). It is known for its use of the Hindley-Milner type inference algorithm, which can automatically infer the types of most expressions without requiring explicit type annotations.
ML is often referred to as an impure functional language, because it does not encapsulate side-effects, unlike purely functional programming languages such as {\haskell}.

Features of ML include a call-by-value evaluation strategy, first class functions, automatic memory management through garbage collection, parametric polymorphism, static typing, type inference, algebraic data types, pattern matching, and exception handling.

Unlike Haskell, ML uses eager evaluation, which means that all subexpressions are always evaluated. However, lazy evaluation can be achieved through the use of closures. Thus one can create and use infinite streams as in Haskell, however, their expression is comparatively indirect.

Today there are several languages in the ML family; the two major dialects are Standard ML (SML) and Caml, but others exist, including F\# -- a language which Microsoft supports for their .NET platform. Ideas from ML have influenced numerous other languages, like
{\mllessequal} \cite{mllessequal}, Haskell, Cyclone, and Nemerle.

ML's strengths are mostly applied in language design and manipulation (compilers, analyzers, theorem provers \cite{isabelle,isabelle-hol-tutorial}), but it is a general-purpose language also used in bioinformatics, financial systems, and applications including a genealogical database, a peer-to-peer client/server program, etc.

\begin{enumerate}
	\item 
\url{http://en.wikipedia.org/wiki/ML_%28programming_language%29}
\end{enumerate}

\subsubsection{{\scheme}}

{\scheme} is one of the two main dialects of the programming language Lisp. Unlike Common Lisp, the other main dialect, {\scheme} follows a minimalist design philosophy specifying a small standard core with powerful tools for language extension. Its compactness and elegance have made it popular with educators, language designers, programmers, implementors, and hobbyists, and this diverse appeal is seen as both a strength and, because of the diversity of its constituencies and the wide divergence between implementations, one of its weaknesses.

{\scheme} was developed at the MIT AI Lab by Guy L. Steele and Gerald Jay Sussman who introduced it to the academic world via a series of memos, now referred to as the Lambda Papers, over the period 1975-1980. {\scheme} had a significant influence on the effort that led to the development of its sister, Common LISP.

\begin{enumerate}
	\item 
\url{http://en.wikipedia.org/wiki/Scheme_%28programming_language%29}
\end{enumerate}

\subsubsection{{\cpp}}

{\cpp} is a statically typed, free-form, multi-paradigm, compiled, general-purpose programming language. It is regarded as a middle-level language, as it comprises a combination of both high-level and low-level language features. It was developed by Bjarne Stroustrup starting in 1979 at Bell Labs as an enhancement to the {\C} programming language and originally named ``C with Classes''. It was renamed {\cpp} in 1983.

{\cpp} is widely used in the software industry, and remains one of the most popular languages ever created. Some of its application domains include systems software, application software, device drivers, embedded software, high-performance server and client applications, and entertainment software such as video games. Several groups provide both free and proprietary {\cpp} compiler software, including the GNU Project, Microsoft, Intel, Borland and others.

The language began as enhancements to C, first adding classes, then virtual functions, operator overloading, multiple inheritance, templates, and exception handling among other features. 

Stroustrup began work on ``C with Classes'' in 1979. The idea of creating a new language originated from Stroustrup's experience in programming for his Ph.D. thesis. Stroustrup found that Simula had features that were very helpful for large software development, but the language was too slow for practical use, while BCPL (the ancestor of C) was fast but too low-level to be suitable for large software development. Remembering his Ph.D. experience, Stroustrup set out to enhance the C language with Simula-like features. C was chosen because it was general-purpose, fast, portable and widely used. Besides C and Simula, some other languages that inspired him were ALGOL 68, Ada, CLU and ML. At first, the class, derived class, strong type checking, inlining, and default argument features were added to C via Cfront. The first commercial release occurred in October 1985.[3]
In 1983, the name of the language was changed from C with Classes to {\cpp} (++ being the increment operator in C and {\cpp}). New features were added including virtual functions, function name and operator overloading, references, constants, user-controlled free-store memory control, improved type checking, and BCPL style single-line comments with two forward slashes (//). In 1985, the first edition of The {\cpp} Programming Language was released, providing an important reference to the language, since there was not yet an official standard. Release 2.0 of {\cpp} came in 1989. New features included multiple inheritance, abstract classes, static member functions, const member functions, and protected members. In 1990, The Annotated {\cpp} Reference Manual was published. This work became the basis for the future standard. Late addition of features included templates, exceptions, namespaces, new casts, and a Boolean type.

As the {\cpp} language evolved, a standard library also evolved with it. The first addition to the {\cpp} standard library was the stream I/O library which provided facilities to replace the traditional C functions such as \api{printf} and \api{scanf}. Later, among the most significant additions to the standard library, was the Standard Template Library.
{\cpp} continues to be used and is one of the preferred programming languages to develop professional applications. The language has gone from being mostly Western to attracting programmers from all over the world.

Bjarne Stroustrup describes some rules that he uses for the design of {\cpp}:

\begin{itemize}
	\item 
{\cpp} is designed to be a statically typed, general-purpose language that is as efficient and portable as C
	\item 
{\cpp} is designed to directly and comprehensively support multiple programming styles (procedural programming, data abstraction, object-oriented programming, and generic programming)
	\item 
{\cpp} is designed to give the programmer choice, even if this makes it possible for the programmer to choose incorrectly
	\item 
{\cpp} is designed to be as compatible with C as possible, therefore providing a smooth transition from C
	\item 
{\cpp} avoids features that are platform specific or not general purpose
	\item 
{\cpp} does not incur overhead for features that are not used (the ``zero-overhead principle'')
	\item 
{\cpp} is designed to function without a sophisticated programming environment
Stroustrup also mentions that {\cpp} was always intended to make programming more fun and that many of the double meanings in the language are intentional, such as the name of the language.
\end{itemize}

\begin{enumerate}
	\item 
\url{http://en.wikipedia.org/wiki/C%2B%2B}
\end{enumerate}

\subsubsection{Ada}
\index{Ada}

Ada is a structured, statically typed, imperative, wide-spectrum, and object-oriented high-level computer programming language, extended from Pascal and other languages. It was originally designed by a team led by Jean Ichbiah of CII Honeywell Bull under contract to the United States Department of Defense (DoD) from 1977 to 1983 to supersede the hundreds of programming languages then used by the DoD. Ada is strongly typed and compilers are validated for reliability in mission-critical applications, such as avionics software. Ada was named after Ada Lovelace (1815–1852), who is often credited as being the first computer programmer.

\begin{enumerate}
	\item 
\url{http://en.wikipedia.org/wiki/Ada_%28programming_language%29}
\end{enumerate}

\subsubsection{Perl}

{\perl} is a high-level, general-purpose, interpreted, dynamic programming language. Perl was originally developed by Larry Wall, a linguist working as a systems administrator for NASA, in 1987, as a general-purpose Unix scripting language to make report processing easier. Since then, it has undergone many changes and revisions and become widely popular amongst programmers. Larry Wall continues to oversee development of the core language, and its upcoming version, Perl 6.
Perl borrows features from other programming languages including C, shell scripting (sh), AWK, and \tool{sed} \cite{sed}. The language provides powerful text processing facilities without the arbitrary data length limits of many contemporary Unix tools, facilitating easy manipulation of text files. It is also used for graphics programming, system administration, network programming, applications that require database access and CGI programming on the Web. 

\begin{enumerate}
	\item 
\url{http://en.wikipedia.org/wiki/Perl}
\end{enumerate}

\subsubsection{{\python}}

{\python} is a general-purpose high-level programming language. Its design philosophy emphasizes code readability. {\python} claims to ``[combine] remarkable power with very clear syntax'', and its standard library is large and comprehensive. Its use of indentation for block delimiters is unusual among popular programming languages.

{\python} supports multiple programming paradigms (primarily object oriented, imperative, and functional) and features a fully dynamic type system and automatic memory management, similar to {\perl}, Ruby, {\scheme}, and Tcl. Like other dynamic languages, {\python} is often used as a scripting language.

The language has an open, community-based development model managed by the non-profit {\python} Software Foundation, which maintains the de facto definition of the language in CPython, the reference implementation. 

{\python} is a multi-paradigm programming language. Rather than forcing programmers to adopt a particular style of programming, it permits several styles: object-oriented programming and structured programming are fully supported, and there are a number of language features which support functional programming and aspect-oriented programming (including by metaprogramming and by magic methods). Many other paradigms are supported using extensions, such as pyDBC and Contracts for {\python} which allow Design by Contract.

{\python} is often used as a scripting language for web applications, and has seen extensive use in the information security industry. {\python} has been successfully embedded in a number of software products as a scripting language, including in finite element method software such as Abaqus, 3D animation packages such as Maya \cite{maya}, MotionBuilder, Softimage, Cinema 4D, BodyPaint 3D, modo, and Blender \cite{blender}, and 2D imaging programs like GIMP, Inkscape, Scribus, and Paint Shop Pro. It has even been used in several videogames.

For many operating systems, {\python} is a standard component; it ships with most Linux distributions, with NetBSD, and OpenBSD, and with \macos{X}. 

Among the users of {\python} are YouTube and the original BitTorrent client. Large organizations that make use of {\python} include Google, Yahoo!, CERN and NASA. 

\begin{enumerate}
	\item 
\url{http://en.wikipedia.org/wiki/Python_%28programming_language%29}
\end{enumerate}

\subsubsection{Haskell}

{\haskell} is a standardized, general-purpose purely functional programming language, with non-strict semantics and strong static typing. It is named after logician Haskell Curry. 

Haskell is a purely functional language, which means that in general, functions in Haskell do not have side effects. There is a distinct type for representing side effects, orthogonal to the type of functions. A pure function may return a side effect which is subsequently executed, modeling the impure functions of other languages. Haskell has a non-strict semantics. Most implementations of Haskell use lazy evaluation. Haskell has a strong, static, type system based on Hindley-Milner type inference. Haskell's principal innovation in this area is to add type classes, which were originally conceived as a principled way to add overloading to the language, but have since found many more uses

Following the release of Miranda by Research Software Ltd, in 1985, interest in lazy functional languages grew. By 1987, more than a dozen non-strict, purely functional programming languages existed. Of these, Miranda was the most widely used, but was not in the public domain. At the conference on Functional Programming Languages and Computer Architecture (FPCA '87) in Portland, Oregon, a meeting was held during which participants formed a strong consensus that a committee should be formed to define an open standard for such languages. The committee's purpose was to consolidate the existing functional languages into a common one that would serve as a basis for future research in functional-language design. The first version of Haskell ("Haskell 1.0") was defined in 1990. The committee's efforts resulted in a series of language definitions. In late 1997, the series culminated in Haskell 98, intended to specify a stable, minimal, portable version of the language and an accompanying standard library for teaching, and as a base for future extensions. 

\begin{enumerate}
	\item 
\url{http://en.wikipedia.org/wiki/Haskell_%28programming_language%29}
\end{enumerate}

\subsubsection{Java}
\index{Java}

Java is a programming language originally developed by James Gosling at Sun Microsystems and released in 1995 as a core component of Sun Microsystems' Java platform. The language derives much of its syntax from {\C} and {\cpp} but has a simpler object model and fewer low-level facilities. Java applications are typically compiled to bytecode (class file) that can run on any Java Virtual Machine (JVM) regardless of computer architecture. Java is general-purpose, concurrent, class-based, and object-oriented, and is specifically designed to have as few implementation dependencies as possible. It is intended to let application developers ``write once, run anywhere''.

There were five primary goals in the creation of the Java language:

\begin{itemize}
	\item 
It should be ``simple, object oriented, and familiar''.
	\item 
It should be ``robust and secure''.
	\item 
It should be ``architecture neutral and portable''.
	\item 
It should execute with ``high performance''.
	\item 
It should be ``interpreted, threaded, and dynamic''.
\end{itemize}

Programs written in Java have a reputation for being slower and requiring more memory than those written in some other languages. However, Java programs' execution speed improved significantly with the introduction of Just-in-time compilation in 1997/1998 for Java 1.1, the addition of language features supporting better code analysis, and optimizations in the Java Virtual Machine itself, such as HotSpot becoming the default for Sun's JVM in 2000.

{\java} uses an automatic garbage collector to manage memory in the object lifecycle. The programmer determines when objects are created, and the Java runtime is responsible for recovering the memory once objects are no longer in use. Once no references to an object remain, the unreachable memory becomes eligible to be freed automatically by the garbage collector. Something similar to a memory leak may still occur if a programmer's code holds a reference to an object that is no longer needed, typically when objects that are no longer needed are stored in containers that are still in use. If methods for a nonexistent object are called, a ``null pointer exception'' is thrown.

One of the ideas behind Java's automatic memory management model is that programmers be spared the burden of having to perform manual memory management. In some languages memory for the creation of objects is implicitly allocated on the stack, or explicitly allocated and deallocated from the heap. Either way, the responsibility of managing memory resides with the programmer. If the program does not deallocate an object, a memory leak occurs. If the program attempts to access or deallocate memory that has already been deallocated, the result is undefined and difficult to predict, and the program is likely to become unstable and/or crash. This can be partially remedied by the use of smart pointers, but these add overhead and complexity. Note that garbage collection does not prevent 'logical' memory leaks, i.e. those where the memory is still referenced but never used.
Garbage collection may happen at any time. Ideally, it will occur when a program is idle. It is guaranteed to be triggered if there is insufficient free memory on the heap to allocate a new object; this can cause a program to stall momentarily. Explicit memory management is not possible in Java.

Java does not support {\C}/{\cpp} style pointer arithmetic, where object addresses and unsigned integers (usually long integers) can be used interchangeably. This allows the garbage collector to relocate referenced objects, and ensures type safety and security.
As in {\cpp} and some other object-oriented languages, variables of Java's primitive data types are not objects. This was a conscious decision by Java's designers for performance reasons. Because of this, Java was not considered to be a pure object-oriented programming language. However, as of Java 5.0, autoboxing enables programmers to proceed as if primitive types are instances of their wrapper classes.

\begin{enumerate}
	\item 
\url{http://en.wikipedia.org/wiki/Java_%28programming_language%29}
\end{enumerate}

\section{References}

\begin{enumerate}
	\item 
\url{http://en.wikipedia.org/wiki/Timeline_of_programming_languages}

	\item 
\url{http://merd.sourceforge.net/pixel/language-study/diagram.html}

	\item 
\url{http://www.levenez.com/lang/}

	\item 
\url{http://oreilly.com/pub/a/oreilly/news/languageposter_0504.html}

	\item 
\url{http://www.digibarn.com/collections/posters/tongues/ComputerLanguagesChart-med.png}
\end{enumerate}

\chapter{Programming Paradigms}
\label{chapt:programming-paradigms}

\section{Introduction}

A programming paradigm is a fundamental style of computer programming. (Compare with a methodology, which is a style of solving specific software engineering problems.) Paradigms differ in the concepts and abstractions used to represent the elements of a program (such as objects, functions, variables, constraints, etc.) and the steps that compose a computation (assignation, evaluation, continuations, data flows, etc.).

A programming paradigm can be understood as an abstraction of a computer system, for example the
von Neumann model
used in traditional sequential computers. For parallel computing, there are many possible models typically reflecting different ways processors can be interconnected. The most common are based on shared memory, distributed memory with message passing, or a hybrid of the two. In object-oriented programming, programmers can think of a program as a collection of interacting objects, while in functional programming a program can be thought of as a sequence of stateless function evaluations. When programming computers or systems with many processors, process-oriented programming allows programmers to think about applications as sets of concurrent processes acting upon logically shared data structures. 

Just as different groups in software engineering advocate different methodologies, different programming languages advocate different programming paradigms. Some languages are designed to support one particular paradigm (Smalltalk supports object-oriented programming, Haskell supports functional programming), while other programming languages support multiple paradigms (such as Object Pascal, {\cpp}, {\csharp}, Visual Basic, Common Lisp, {\scheme}, {\perl}, {\python}, Ruby, Oz and F Sharp).

A multi-paradigm programming language is a programming language that supports more than one programming paradigm. The design goal of such languages is to allow programmers to use the best tool for a job, admitting that no one paradigm solves all problems in the easiest or most efficient way. An example is 
Oz,
which has subsets that are a logic language (Oz descends from logic programming), a functional language, an object-oriented language, a dataflow concurrent language, and more. Oz was designed over a ten-year period to combine in a harmonious way concepts that are traditionally associated with different programming paradigms.

\section{History}

\subsection{Low-level: binary, assembly}
\index{Assembly}
\index{low-level}

Initially, computers were hard-wired\index{hard-wired} or soft-wired\index{hard-wired} and then later programmed
using binary code that represented control sequences fed to the computer CPU.
This was difficult and error-prone. Programs written in binary are said to be written
in machine code, which is a very low-level programming paradigm.
Hard-wired, soft-wired, and binary programming are considered first generation languages. 

To make programming easier, assembly languages were developed. These replaced machine
code functions with mnemonics and memory addresses with symbolic labels.
Assembly language programming is considered a low-level paradigm although
it is a ``second generation'' paradigm. Even assembly languages of the 1960s actually
supported library \api{COPY} and quite sophisticated conditional macro generation and pre-processing
capabilities. They also supported modular programming features such as \api{CALL} (subroutines),
external variables and common sections (globals), enabling significant code re-use and
isolation from hardware specifics via use of logical operators as \api{READ}/\api{WRITE}/\api{GET}/\api{PUT}.
Assembly was, and still is, used for time critical systems and frequently in embedded systems\index{embedded systems}.
Assembly languages are considered second generation languages.

\begin{enumerate}
	\item 
\url{http://en.wikipedia.org/wiki/Machine_code}

	\item 
\url{http://en.wikipedia.org/wiki/Low-level_programming_language}

	\item 
\url{http://en.wikipedia.org/wiki/Assembly_programming}
\end{enumerate}

\subsection{Procedural programming}

Procedural programming can sometimes be used as a synonym for imperative programming (specifying the steps the program must take to reach the desired state), but can also refer (as in this article) to a programming paradigm, derived from structured programming, based upon the concept of the procedure call. Procedures, also known as routines, subroutines, methods, or functions (not to be confused with mathematical functions, but similar to those used in functional programming) simply contain a series of computational steps to be carried out. Any given procedure might be called at any point during a program's execution, including by other procedures or itself. A procedural programming language provides a programmer a means to define precisely each step in the performance of a task. The programmer knows what is to be accomplished and provides through the language step-by-step instructions on how the task is to be done. Using a procedural language, the programmer specifies language statements to perform a sequence of algorithmic steps. Procedural programming is often a better choice than simple sequential or unstructured programming in many situations which involve moderate complexity or which require significant ease of maintainability. 

Possible benefits:

\begin{itemize}
	\item 
 The ability to re-use the same code at different places in the program without copying it.
	\item 
 An easier way to keep track of program flow than a collection of ``GOTO'' or ``JUMP'' statements (which can turn a large, complicated program into spaghetti code).
	\item 
 The ability to be strongly modular or structured.
\end{itemize}

Thus, procedural programming allows for '''modularity''', which is generally desirable, especially in large, complicated programs. Inputs are usually specified syntactically in the form of arguments and the outputs delivered as return values. '''Scoping''' is another technique that helps keep procedures strongly modular. It prevents a procedure from accessing the variables of other procedures (and vice-versa), including previous instances of itself such as in recursion, without explicit authorization. Because of the ability to specify a simple interface, to be self-contained, and to be reused, procedures are a convenient vehicle for making pieces of code written by different people or different groups, including through programming '''libraries'''. 

The focus of procedural programming is to break down a programming task into a collection of variables, data structures, and subroutines, whereas in object-oriented programming it is to break down a programming task into objects with each ``object'' encapsulating its own data and methods (subroutines). The most important distinction is whereas procedural programming uses procedures to operate on data structures, object-oriented programming bundles the two together so an ``object'' operates on its ``own'' data structure.

The earliest imperative languages were the machine languages of the original computers. In these languages, instructions were very simple, which made hardware implementation easier, but hindered the creation of complex programs. FORTRAN, developed by John Backus at IBM starting in 1954, was the first major programming language to remove the obstacles presented by machine code in the creation of complex programs. FORTRAN was a compiled language that allowed named variables, complex expressions, subprograms, and many other features now common in imperative languages. The next two decades saw the development of a number of other major high-level imperative programming languages. In the late 1950s and 1960s, ALGOL was developed in order to allow mathematical algorithms to be more easily expressed, and even served as the operating system's target language for some computers. COBOL (1960) and BASIC (1964) were both attempts to make programming syntax look more like English. In the 1970s, Pascal was developed by Niklaus Wirth, and C was created by Dennis Ritchie while he was working at Bell Laboratories. For the needs of the United States Department of Defense, Jean Ichbiah and a team at Honeywell began designing Ada in 1978.

\begin{enumerate}
	\item 
\url{http://en.wikipedia.org/wiki/Procedural_programming}

	\item 
\url{http://en.wikipedia.org/wiki/Imperative_programming}
\end{enumerate}

\subsection{Object-oriented programming}

Object-oriented programming (OOP) is a programming paradigm that uses ``objects'' – data structures consisting of data fields and methods together with their interactions – to design applications and computer programs. Programming techniques may include features such as data abstraction, encapsulation, modularity, polymorphism, and inheritance. Though it was invented with the creation of the [[Wikipedia:Simula|Simula]] language in 1965, it was not commonly used in mainstream software application development until the early 1990s. Many modern programming languages now support OOP.

There is still some controversy by notable programmers such as
Alexander Stepanov,
Richard Stallman
and others, concerning the efficacy of the OOP paradigm versus the procedural paradigm. The necessity of every object to have associative methods leads some skeptics to associate OOP with software bloat. Polymorphism was developed as one attempt to resolve this dilemma.

Object-oriented programming is characterized by a group of inter-related fundamental concepts: 

\begin{itemize}
	\item 
``Class'':
Defines the abstract characteristics of a thing (object), including the thing's characteristics (its attributes, fields or properties) and the thing's behaviors (the things it can do, or methods, operations or features). One might say that a class is a blueprint or factory that describes the nature of something. For example, the class Dog would consist of traits shared by all dogs, such as breed and fur color (characteristics), and the ability to bark and sit (behaviors). Classes provide modularity and structure in an object-oriented computer program. A class should typically be recognizable to a non-programmer familiar with the problem domain, meaning that the characteristics of the class should make sense in context. Also, the code for a class should be relatively self-contained (generally using encapsulation). Collectively, the properties and methods defined by a class are called members.
	\item 
``Object'':
An individual of a class. The class Dog defines all possible dogs by listing the characteristics and behaviors they can have; the object Lassie is one particular dog, with particular versions of the characteristics. A Dog has fur; Lassie has brown-and-white fur. One can have an instance of a class or a particular object. The instance is the actual object created at runtime. In programmer jargon, the Lassie object is an instance of the Dog class. The set of values of the attributes of a particular object is called its state. The object consists of state and the behaviour that's defined in the object's class.
	\item 
``Method'':
An object's abilities. In language, methods (sometimes referred to as ``functions") are verbs. Lassie, being a Dog, has the ability to bark. So \api{bark()} is one of Lassie's methods. She may have other methods as well, for example sit() or eat() or walk() or saveTimmy(). Within the program, using a method usually affects only one particular object; all Dogs can bark, but you need only one particular dog to do the barking.
	\item 
``Message passing'':
The process by which an object sends data to another object or asks the other object to invoke a method. Also known to some programming languages as interfacing. For example, the object called Breeder may tell the Lassie object to sit by passing a ``sit'' message which invokes Lassie's ``sit'' method. The syntax varies between languages, for example: [Lassie sit] in Objective-C. In {\java}, code-level message passing corresponds to ``method calling''. Some dynamic languages use double-dispatch or multi-dispatch to find and pass messages.
	\item 
``Inheritance'':
``Subclasses'' are more specialized versions of a class, which inherit attributes and behaviors from their parent classes, and can introduce their own. For example, the class Dog might have sub-classes called Collie, Chihuahua, and GoldenRetriever. In this case, Lassie would be an instance of the Collie subclass. Suppose the Dog class defines a method called \api{bark()} and a property called furColor. Each of its sub-classes (Collie, Chihuahua, and GoldenRetriever) will inherit these members, meaning that the programmer only needs to write the code for them once. Each subclass can alter its inherited traits. For example, the Collie subclass might specify that the default furColor for a collie is brown-and-white. The Chihuahua subclass might specify that the \api{bark()} method produces a high pitch by default. Subclasses can also add new members. The Chihuahua subclass could add a method called tremble(). So an individual chihuahua instance would use a high-pitched \api{bark()} from the Chihuahua subclass, which in turn inherited the usual \api{bark()} from Dog. The chihuahua object would also have the tremble() method, but Lassie would not, because she is a Collie, not a Chihuahua. In fact, inheritance is an ``a... is a'' relationship between classes, while instantiation is an ``is a'' relationship between an object and a class: a Collie is a Dog (``a... is a''), but Lassie is a Collie (``is a''). Thus, the object named Lassie has the methods from both classes Collie and Dog. Multiple inheritance is inheritance from more than one ancestor class, neither of these ancestors being an ancestor of the other. For example, independent classes could define Dogs and Cats, and a Chimera object could be created from these two which inherits all the (multiple) behavior of cats and dogs. This is not always supported, as it can be hard both to implement and to use well.
	\item 
``Abstraction'':
Abstraction is simplifying complex reality by modeling classes appropriate to the problem, and working at the most appropriate level of inheritance for a given aspect of the problem. For example, Lassie the Dog may be treated as a Dog much of the time, a Collie when necessary to access Collie-specific attributes or behaviors, and as an Animal (perhaps the parent class of Dog) when counting Timmy's pets. Abstraction is also achieved through Composition. For example, a class Car would be made up of an Engine, Gearbox, Steering objects, and many more components. To build the Car class, one does not need to know how the different components work internally, but only how to interface with them, i.e., send messages to them, receive messages from them, and perhaps make the different objects composing the class interact with each other.
	\item 
``Encapsulation'':
Encapsulation conceals the functional details of a class from objects that send messages to it. For example, the Dog class has a \api{bark()} method. The code for the \api{bark()} method defines exactly how a bark happens (e.g., by \api{inhale()} and then exhale(), at a particular pitch and volume). Timmy, Lassie's friend, however, does not need to know exactly how she barks. Encapsulation is achieved by specifying which classes may use the members of an object. The result is that each object exposes to any class a certain interface -- those members accessible to that class. The reason for encapsulation is to prevent clients of an interface from depending on those parts of the implementation that are likely to change in the future, thereby allowing those changes to be made more easily, that is, without changes to clients. Members are often specified as public, protected or private, determining whether they are available to all classes, sub-classes or only the defining class. Some languages go further: {\java} uses the default access modifier to restrict access also to classes in the same package, {\csharp} and VB.NET reserve some members to classes in the same assembly using keywords internal ({\csharp}) or Friend (VB.NET), and Eiffel and {\cpp} allow one to specify which classes may access any member.
	\item 
``Polymorphism'':
Polymorphism allows the programmer to treat derived class members just like their parent class' members. More precisely, Polymorphism in object-oriented programming is the ability of objects belonging to different data types to respond to method calls of methods of the same name, each one according to an appropriate type-specific behavior. One method, or an operator such as +, -, or *, can be abstractly applied in many different situations. If a Dog is commanded to \api{speak()}, this may elicit a \api{bark()}. However, if a Pig is commanded to \api{speak()}, this may elicit an oink(). They both inherit \api{speak()} from Animal, but their derived class methods override the methods of the parent class; this is '''Overriding Polymorphism'''. '''Overloading Polymorphism''' is the use of one method signature, or one operator such as ``+'', to perform several different functions depending on the implementation. The ``+'' operator, for example, may be used to perform integer addition, float addition, list concatenation, or string concatenation. Any two subclasses of Number, such as Integer and Double, are expected to add together properly in an OOP language. The language must therefore overload the addition operator, ``+'', to work this way. This helps improve code readability. How this is implemented varies from language to language, but most OOP languages support at least some level of overloading polymorphism. Many OOP languages also support '''parametric polymorphism''', where code is written without mention of any specific type and thus can be used transparently with any number of new types. The use of pointers to a superclass type later instantiated to an object of a subclass is a simple yet powerful form of polymorphism, such as used in {\cpp}.
\end{itemize}

The 1980s saw a rapid growth in interest in object-oriented programming. These languages were imperative in style, but added features to support objects. The last two decades of the 20th century saw the development of a considerable number of such programming languages. Smalltalk-80, originally conceived by Alan Kay\index{Alan Kay} in 1969, was released in 1980 by the Xerox Palo Alto Research Center. Drawing from concepts in another object-oriented language -- Simula (which is considered to be the world's first object-oriented programming language, developed in the late 1960s)—Bjarne Stroustrup designed {\cpp}, an object-oriented language based on {\C}. {\cpp} was first implemented in 1985. In the late 1980s and 1990s, the notable imperative languages drawing on object-oriented concepts were {\perl}, released by Larry Wall in 1987; {\python}, released by Guido van Rossum in 1990; {\php}, released by Rasmus Lerdorf in 1994; {\java}, first released by Sun Microsystems in 1994 and Ruby, released in 1995 by Yukihiro Matsumoto. 

\begin{enumerate}
	\item 
\url{http://en.wikipedia.org/wiki/Object-oriented_programming}

	\item 
\url{http://en.wikipedia.org/wiki/Class_(computer_science)}

	\item 
\url{http://en.wikipedia.org/wiki/Instance_(programming)}

	\item 
\url{http://en.wikipedia.org/wiki/Method_(computer_science)}

	\item 
\url{http://en.wikipedia.org/wiki/Message_passing}

	\item 
\url{http://en.wikipedia.org/wiki/Inheritance_(object-oriented_programming)}

	\item 
\url{http://en.wikipedia.org/wiki/Abstraction_(computer_science)}

	\item 
\url{http://en.wikipedia.org/wiki/Encapsulation_(computer_science)}

	\item 
\url{http://en.wikipedia.org/wiki/Polymorphism_in_object-oriented_programming}
\end{enumerate}

\subsubsection{Aspect-oriented programming}
\index{Aspect-oriented programming}
\index{AOP}

Aspect-oriented programming (AOP) is a programming paradigm in which secondary or supporting functions are isolated from the main program's business logic. It aims to increase modularity by allowing the separation of cross-cutting concerns, forming a basis for aspect-oriented software development.

Aspect-oriented programming entails breaking down a program into distinct parts (so-called concerns, cohesive areas of functionality). All programming paradigms support some level of grouping and encapsulation of concerns into separate, independent entities by providing abstractions (e.g. procedures, modules, classes, methods) that can be used for implementing, abstracting and composing these concerns. But some concerns defy these forms of implementation and are called crosscutting concerns because they ``cut across'' multiple abstractions in a program. Logging is a common example of a crosscutting concern because a logging strategy necessarily affects every single logged part of the system. Logging thereby crosscuts all logged classes and methods.

All AOP implementations have some crosscutting expressions that encapsulate each concern in one place. The difference between implementations lies in the power, safety, and usability of the constructs provided. For example, interceptors that specify the methods to intercept express a limited form of crosscutting, without much support for type-safety or debugging. {\aspectj} \cite{aspectj} has a number of such expressions and encapsulates them in a special class type, called an \api{aspect}. For example, an aspect can alter the behavior of the base code (the non-aspect part of a program) by applying advice (additional behavior) at various join points (points in a program) specified in a quantification or query called a pointcut (that detects whether a given join point matches). An aspect can also make binary-compatible structural changes to other classes, like adding members or parents.

The following are some standard terminology used in Aspect-oriented programming:

\begin{itemize}
	\item 
 {\bf Cross-cutting concerns}: Even though most classes in an OO model will perform a single, specific function, they often share common, secondary requirements with other classes. For example, we may want to add logging to classes within the data-access layer and also to classes in the UI layer whenever a thread enters or exits a method. Even though the primary functionality of each class is very different, the code needed to perform the secondary functionality is often identical.
	\item 
 {\bf Advice}: This is the additional code that you want to apply to your existing model. In our example, this is the logging code that we want to apply whenever the thread enters or exits a method.
	\item 
 {\bf Pointcut}: This is the term given to the point of execution in the application at which cross-cutting concern needs to be applied. In our example, a pointcut is reached when the thread enters a method, and another pointcut is reached when the thread exits the method.
	\item 
 {\bf Aspect}: The combination of the pointcut and the advice is termed an aspect. 
\end{itemize}

Programmers need to be able to read code and understand what is happening in order to prevent errors. Even with proper education, understanding crosscutting concerns can be difficult without proper support for visualizing both static structure and the dynamic flow of a program. Visualizing crosscutting concerns is just beginning to be supported in IDEs, as is support for aspect code assist and refactoring. Given the power of AOP, if a programmer makes a logical mistake in expressing crosscutting, it can lead to widespread program failure. Conversely, another programmer may change the join points in a program – e.g., by renaming or moving methods – in ways that were not anticipated by the aspect writer, with unintended consequences. One advantage of modularizing crosscutting concerns is enabling one programmer to affect the entire system easily; as a result, such problems present as a conflict over responsibility between two or more developers for a given failure. However, the solution for these problems can be much easier in the presence of AOP, since only the aspect need be changed, whereas the corresponding problems without AOP can be much more spread out.

\begin{enumerate}
	\item 
\url{http://en.wikipedia.org/wiki/Aspect-oriented_programming}

	\item 
\url{http://en.wikipedia.org/wiki/Aspect-Oriented_Software_Development}

	\item 
\url{http://www.eclipse.org/aspectj/}

	\item 
\url{http://www.sable.mcgill.ca/publications/techreports/sable-tr-2004-2.pdf}
\end{enumerate}

\subsubsection{Reflection-oriented programming}

Reflection-oriented programming, or reflective programming, is a functional extension to the object-oriented programming paradigm. Reflection-oriented programming includes self-examination, self-modification, and self-replication. However, the emphasis of the reflection-oriented paradigm is dynamic program modification, which can be determined and executed at runtime. Some imperative approaches, such as procedural and object-oriented programming paradigms, specify that there is an exact predetermined sequence of operations with which to process data. The reflection-oriented programming paradigm, however, adds that program instructions can be modified dynamically at runtime and invoked in their modified state. That is, the program architecture itself can be decided at runtime based upon the data, services, and specific operations that are applicable at runtime.

Programming sequences can be classified in one of two ways, atomic or compound. Atomic operations are those that can be viewed as completing in a single, logical step, such as the addition of two numbers. Compound operations are those that require a series of multiple atomic operations.

A compound statement, in classic procedural or object-oriented programming, can lose its structure once it is compiled. The reflective programming paradigm introduces the concept of meta-information, which keeps knowledge of program structure. Meta-information stores information such as the name of the contained methods, the name of the class, the name of parent classes, and/or what the compound statement is supposed to do. Using this stored information, as an object is consumed (processed), it can be reflected upon to find out the operations that it supports. The operation that issues in the required state via the desired state transition can be chosen at run-time without hard-coding it.

Reflection can be used for observing and/or modifying program execution at runtime. A reflection-oriented program component can monitor the execution of an enclosure of code and can modify itself according to a desired goal related to that enclosure. This is typically accomplished by dynamically assigning program code at runtime \cite{java-reflection}.

Reflection can also be used to adapt a given program to different situations dynamically. For example, consider an application that uses two different classes X and Y interchangeably to perform similar operations. Without reflection-oriented programming, the application might be hard-coded to call method names of class X and class Y. However, using the reflection-oriented programming paradigm, the application could be designed and written to utilize reflection in order to invoke methods in classes X and Y without hard-coding method names. Reflection-oriented programming almost always requires additional knowledge, framework, relational mapping, and object relevance in order to take advantage of more generic code execution. Hard-coding can be avoided to the extent that reflection-oriented programming is used. Reflection is also a key strategy for metaprogramming.

A language supporting reflection provides a number of features available at runtime that would otherwise be very obscure or impossible to accomplish in a lower-level language. Some of these features are the abilities to:

\begin{itemize}
	\item 
 Discover and modify source code constructions (such as code blocks, classes, methods, protocols, etc.) as a first-class object at runtime.
	\item 
 Convert a string matching the symbolic name of a class or function into a reference to or invocation of that class or function.
	\item 
 Evaluate a string as if it were a source code statement at runtime.
	\item 
 Create a new interpreter for the language's bytecode to give a new meaning or purpose for a programming construct.
\end{itemize}

These features can be implemented in different ways. Compiled languages rely on their runtime system to provide information about the source code. A compiled Objective-C executable, for example, records the names of all methods in a block of the executable, providing a table to correspond these with the underlying methods (or selectors for these methods) compiled into the program. In a compiled language that supports runtime creation of functions, such as Common Lisp, the runtime environment must include a compiler or an interpreter. Reflection can be implemented for languages not having built-in reflection facilities by using a program transformation system to define automated source code changes.

\begin{enumerate}
	\item 
\url{http://en.wikipedia.org/wiki/Reflection_%28computer_science%29}
\end{enumerate}

\subsection{Declarative programming}

Declarative programming is a programming paradigm that expresses the logic of a computation without describing its control flow. Many languages applying this style attempt to minimize or eliminate side effects by describing what the program should accomplish, rather than describing how to go about accomplishing it. This is in contrast with imperative programming, which requires an explicitly provided algorithm. In declarative programming, the program is structured as a collection of properties to find in the expected result, not as a procedure to follow. For example, given a database or a set of rules, the computer tries to find a solution matching all the desired properties. Common declarative languages include those of regular expressions, logic programming, and functional programming. The archetypical example of a declarative language is the fourth generation language SQL.
 
Declarative programming is often defined as any style of programming that is not imperative. A number of other common definitions exist that attempt to give the term a definition other than simply contrasting it with imperative programming. For example:

\begin{itemize}
	\item 
 A program that describes what computation should be performed and not how to compute it
	\item 
 Any programming language that lacks side effects (or more specifically, is referentially transparent)
	\item 
 A language with a clear correspondence to mathematical logic.
\end{itemize}

\subsubsection{Functional programming}

Functional programming is a programming paradigm that treats computation as the evaluation of mathematical functions and avoids state changes and mutable data. It emphasizes the application of functions, in contrast to the imperative programming style, which emphasizes changes in state. Functional programming has its roots in the lambda calculus, a formal system developed in the 1930s to investigate function definition, function application, and recursion. Many functional programming languages can be viewed as elaborations on the lambda calculus.

In practice, the difference between a mathematical function and the notion of a ``function'' used in imperative programming is that imperative functions can have side effects, changing the value of already calculated variables. Because of this they lack referential transparency, i.e. the same language expression can result in different values at different times depending on the state of the executing program. Conversely, in functional code, the output value of a function depends only on the arguments that are input to the function, so calling a function f twice with the same value for an argument x will produce the same result f(x) both times. Eliminating side-effects can make it much easier to understand and predict the behavior of a program, which is one of the key motivations for the development of functional programming.

Functional programming languages, especially purely functional ones, have largely been emphasized in academia rather than in commercial software development. However, prominent functional programming languages such as {\scheme}, Erlang, OCaml, and {\haskell}, have been used in industrial and commercial applications by a wide variety of organizations. Functional programming also finds use in industry through domain-specific programming languages like R (statistics), Mathematica (symbolic math), J and K (financial analysis), and XSLT (XML). Widespread declarative domain specific languages like SQL and Lex/Yacc, use some elements of functional programming, especially in eschewing mutable values. Spreadsheets can also be viewed as functional programming languages.

Programming in a functional style can also be accomplished in languages that aren't specifically designed for functional programming. For example, the imperative {\perl} programming language has been the subject of a book describing how to apply functional programming concepts. {\java}script, one of the most widely employed languages today, incorporates functional programming capabilities.

John Backus presented the FP programming language in his 1977 Turing Award lecture ``Can Programming Be Liberated From the von Neumann Style? A Functional Style and its Algebra of Programs''. He defines functional programs as being built up in a hierarchical way by means of ``combining forms'' that allow an ``algebra of programs"; in modern language, this means that functional programs follow the principle of compositionality. Backus's paper popularized research into functional programming, though it emphasized function-level programming rather than the lambda-calculus style which has come to be associated with functional programming.

In the 1970s the ML programming language was created by Robin Milner at the University of Edinburgh, and David Turner developed initially the language SASL at the University of St. Andrews and later the language Miranda at the University of Kent. ML eventually developed into several dialects, the most common of which are now Objective Caml and Standard ML. Also in the 1970s, the development of the {\scheme} programming language (a partly-functional dialect of Lisp), as described in the influential ``Lambda Papers' and the 1985 textbook ``Structure and Interpretation of Computer Programs', brought awareness of the power of functional programming to the wider programming-languages community.

The {\haskell} programming language was released in the late 1980s in an attempt to gather together many ideas in functional programming research.

\paragraph{Higher-order functions}

Most functional programming languages use higher-order functions, which are functions that can either take other functions as arguments or return them as results (the differential operator $d/dx$ that produces the derivative of a function f is an example of this in calculus).

Higher-order functions are closely related to first-class functions, in that higher-order functions and first-class functions both allow functions as arguments and results of other functions. The distinction between the two is subtle: ``higher-order'' describes a mathematical concept of functions that operate on other functions, while ``first-class'' is a computer science term that describes programming language entities that have no restriction on their use (thus first-class functions can appear anywhere in the program that other first-class entities like numbers can, including as arguments to other functions and as their return values).

\paragraph{Pure functions}

Purely functional functions (or expressions) have no memory or I/O side effects. They represent a function whose valuation depends only on the value of the parameters it is given. This means that pure functions have several useful properties, many of which can be used to optimize the code:

\begin{itemize}
	\item 
 If the result of a pure expression is not used, it can be removed without affecting other expressions.
	\item 
 If a pure function is called with parameters that cause no side-effects, the result is constant with respect to that parameter list (sometimes called referential transparency), i.e. if the pure function is again called with the same parameters, the same result will be returned (this can enable caching optimizations).
	\item 
 If there is no data dependency between two pure expressions, then their order can be reversed, or they can be performed in parallel and they cannot interfere with one another (in other terms, the evaluation of any pure expression is thread-safe and enables parallel execution).
	\item 
 If the entire language does not allow side-effects, then any evaluation strategy can be used; this gives the compiler freedom to reorder or combine the evaluation of expressions in a program.
\end{itemize}

The notion of pure function is central to code optimization in compilers. While most compilers for imperative programming languages detect pure functions, and perform common-subexpression elimination for pure function calls, they cannot always do this for pre-compiled libraries, which generally do not expose this information, thus preventing optimizations that involve those external functions. Some compilers, such as \tool{gcc} \cite{gcc}, add extra keywords for a programmer to explicitly mark external functions as pure, to enable such optimizations. Fortran 95 allows functions to be designated ``pure'' in order to allow such optimizations.

\paragraph{Recursion}

Iteration in functional languages is usually accomplished via recursion. Recursion may require maintaining a stack, and thus may lead to inefficient memory consumption, but tail recursion can be recognized and optimized by a compiler into the same code used to implement iteration in imperative languages. The {\scheme} programming language standard requires implementations to recognize and optimize tail recursion. Tail recursion optimization can be implemented by transforming the program into continuation passing style during compilation, among other approaches. Common patterns of recursion can be factored out using higher order functions, catamorphisms and anamorphisms, which  ``folds'' and ``unfolds'' a recursive function call nest by using higher-order functions being the most obvious examples. 

\paragraph{Eager vs Lazy evaluation}

Functional languages can be categorized by whether they use strict (eager) or non-strict (lazy) evaluation, concepts that refer to how function arguments are processed when an expression is being evaluated. The technical difference is in the denotational semantics of expressions containing failing or divergent computations. Under strict evaluation, the evaluation of any term containing a failing subterm will itself fail. For example, the expression

\begin{verbatim}
 print length([2+1, 3*2, 1/0, 5-4])
\end{verbatim}

will fail under eager evaluation because of the division by zero in the third element of the list. Under lazy evaluation, the length function will return the value 4 (the length of the list), since evaluating it will not attempt to evaluate the terms making up the list. In brief, eager evaluation always fully evaluates function arguments before invoking the function. Lazy evaluation does not evaluate function arguments unless their values are required to evaluate the function call itself. The usual implementation strategy for lazy evaluation in functional languages is graph reduction, a technique first developed by Chris Wadsworth in 1971. Lazy evaluation is used by default in several pure functional languages, including Miranda, Clean and {\haskell}. 

\paragraph{Type system}

Especially since the development of Hindley-Milner (later Damas-Milner) type inference in the 1970s, functional programming languages have tended to use typed lambda calculus, as opposed to the untyped lambda calculus used in Lisp and its variants (such as {\scheme}). Type inference, or implicit typing, refers to the ability to deduce automatically the type of a value in a programming language. It is a feature present in some strongly statically typed languages. It is often characteristic of — but not limited to — functional programming languages in general. Some languages that include type inference are: Ada, BitC, Boo, {\csharp} 3.0, Cayenne, Clean, Cobra, D, Delphi, Epigram, F\#, {\haskell}, haXe, JavaFX Script, ML, Mythryl, Nemerle, OCaml, Oxygene, Scala, and Visual Basic .NET 9.0. This feature is planned for Fortress, {\cpp}0x and {\perl} 6. The ability to infer types automatically makes many programming tasks easier, leaving the programmer free to omit type annotations while maintaining some level of type safety.

\paragraph{Functional programming in non-functional languages}

It is possible to employ a functional style of programming in languages that are not traditionally considered functional languages. Some non-functional languages have borrowed features such as higher-order functions, and list comprehensions from functional programming languages. This makes it easier to adopt a functional style when using these languages. Functional constructs such as higher-order functions and lazy lists can be obtained in {\cpp} via libraries, such as in {\fcpp} \cite{fcpp1,fcpp2}. In {\C}, function pointers can be used to get some of the effects of higher-order functions. For example the common function map can be implemented using function pointers. In {\csharp} version 3.0 and higher, lambda functions can be employed to write programs in a functional style. Many object-oriented design patterns are expressible in functional programming terms: for example, the Strategy pattern simply dictates use of a higher-order function, and the Visitor pattern roughly corresponds to a Catamorphism, or fold.

\subsubsection{Logic programming}

The logic programming paradigm views computation as automated reasoning over a corpus of knowledge. Facts about the problem domain are expressed as logic formulas, and programs are executed by applying inference rules over them until an answer to the problem is found, or the collection of formulas is proved inconsistent.

Logic programming refers to a paradigm that uses a form of symbolic logic as a programming language, such as first-order logic. Imperative programming and functional programming are essentially about implementing mappings (i.e. functions). Having implemented a mapping M, we can make requests like the following: 

\begin{itemize}
	\item 
  Given $A$, determine the value of $M(A)$
\end{itemize}

A request like this will always have a single answer. Logic programming is based on the notion that a program implements a relation rather than a mapping. Consider two sets of values S and T. R is a relation between S and T if, for every x in S and y in T, R(x,y) is either true or false. If R(x,y) is true, we say that R holds between x and y. Logic programming is about implementing relations. Having implemented a relation, we can make requests like: 

\begin{enumerate}
	\item 
 Given $A$ and $B$, determine whether $R(A,B)$ is true
	\item 
 Given $A$, find all $B$s such that $R(A,B)$ is true
	\item 
 Given $B$, find all $A$s such that $R(A,B)$ is true
	\item 
 Find all $A$s and $B$s for which $R(A,B)$ is true
\end{enumerate}

A request like (1) will have a single answer, yes or no, but the other requests (2)(3) and (4) might have many answers or none. These requests are characteristic of logic programming, and explain why it is potentially higher level than imperative or functional programming. 

Robert Kowalski, one of the fathers of logic programming and automatic theorem proving, introduced the property of logic programming models as

 Logic Programming = Logic Statements + Control Strategy

where Logic Statements consist of a set of rules and facts expressing the relationships between objects. Control Strategy is known as ``resolution'' or the sequence of steps the deduction system chooses to answer a question from the logic system expressed in the program. In other words, it refers to how a logic language computes a response to a question. This indicates that the original set of logic statements represent a knowledge base of the computation, and the deductive system provides the control by which a new logic statement may be derived. This property indicates that in the logic programming system, a programmer needs to be concerned about the logic system in question, but the control system is the evaluation strategy implemented by the program evaluator and can be ignored from his consideration.

\paragraph{Evaluation}

In a logic programming system, there is no indication in a program about how a particular goal might be proved from a given set of predicates. Instead, the evaluation strategy involves the use of ``inference rules'' that allow it to construct a new set of logical statements that are proved to be true from a given set of original logical statements that are already true. This result indicates that the new set of logical statements (the original ones and the inferred ones) can be viewed as representing the potential computation of all logical consequences of a set of original statements. Hence, the essence of a logic program is that from a collection of logical statements, known facts and rules, a desired fact, known as a query or goal, might be proved to be true by applying the inference rules. The two primary operations used by inference rules to derive new facts are ``resolution'' and ``unification''. 

\begin{enumerate}
	\item 
\url{http://en.wikipedia.org/wiki/Logic_programming}
\end{enumerate}


\chapter{Program Evaluation}
\label{chapt:program-evaluation}
\index{Program Evaluation}

In order to be executed, computer programs need to be translated into machine-understandable code.
In order to achieve this, the program to be executed must first be analyzed to determine first
if it is a valid program from the point of view of its lexical conventions (lexical analysis),
syntactical form (syntactical analysis), and semantic validity (semantic analysis). During analysis,
the various components will correctly report errors found, allowing the programmer to understand
the location and nature of the errors found. After the program has been confirmed to be valid,
it is translated (code generation) and possibly optimized (code optimization) before execution.

\section{Program analysis and translation phases}
\index{Program analysis and translation phases}

\subsection{Front end}
\index{Front end}

The front end analyzes the source code to build an internal representation of the program,
called the intermediate representation. It also manages the symbol table\index{symbol table},
a data structure mapping each symbol in the source code to associated information such as location,
type and scope. This is done over several phases, which includes some of the following:

\subsubsection{Lexical analysis}
\index{Lexical analysis}

In computer science, lexical analysis is the process of converting a sequence of characters
into a sequence of tokens. A program or function which performs lexical analysis is called
a lexical analyzer, lexer or scanner. A lexer often exists as a single function,
which is called by the parser. The lexical specification of a programming language
is defined by a set of rules which defines the lexer, which are understood by a lexical
analyzer generator such as \tool{lex} \cite{flex}. The lexical analyzer (either generated automatically by
a tool like \tool{lex}, or hand-crafted) reads in a stream of characters, identifies the lexemes
in the stream, categorizes them into tokens, and outputs a token stream. This is called ``tokenizing.''
If the lexer finds an invalid token, it will report an error.

\subsubsection{Syntactic analysis}
\index{Syntactic analysis}

Syntax analysis involves parsing the token sequence to identify the syntactic structure
of the program. The parser's output is some form of intermediate representation of the
program's structure, typically a parse tree, which replaces the linear sequence of tokens
with a tree structure built according to the rules of a formal grammar, which define the
language's syntax. This is usually done with reference to a context-free grammar,
which recursively defines components that can make up an expression and the order in which
they must appear. The parse tree is often analyzed, augmented, and transformed by
later phases in the compiler. Parsers are written by hand or generated by parser generators,
such as Yacc, Bison \cite{bison,louden97}, or {\javacc} \cite{javacc}.

\subsubsection{Semantic analysis}
\index{Semantic analysis}

Semantic analysis is the phase in which the compiler adds semantic information to the parse
tree and builds the symbol table. This phase performs semantic checks such as type checking 
(checking for type errors), or object binding (associating variable and function references
with their definitions), or definite assignment (requiring all local variables to be initialized
before use), rejecting incorrect programs or issuing warnings. Semantic analysis usually requires
a complete parse tree, meaning that this phase logically follows the parsing phase,
and logically precedes the code generation phase, though it is often possible to fold multiple
phases into one pass over the code in a compiler implementation. Not all rules defining programming
languages can be expressed by context-free grammars alone, for example semantic validity such as
type validity and proper declaration of identifiers. These rules can be formally expressed
with attribute grammars that implement attribute migration across syntax tree nodes when necessary.

\subsection{Back end}
\index{Back end}

The term back end is sometimes confused with code generator because of the overlapped functionality
of generating assembly code. Some literature uses middle end to distinguish the generic analysis
and optimization phases in the back end from the machine-dependent code generators. The main phases
of the back end include the following:

\subsubsection{Analysis}

This is the gathering of program information from the intermediate representation derived
by the front end. Typical analyzes are data flow analysis to build use-define chains,
dependence analysis, alias analysis, pointer analysis, etc. Accurate analysis is the basis
for any compiler optimization. The call graph and control flow graph are usually also built
during the analysis phase.

\subsubsection{Optimization}
\index{Optimization}

The intermediate language representation is transformed into functionally equivalent
but faster (or smaller) forms. Popular optimizations are inline expansion,
dead code elimination, constant propagation, loop transformation, register allocation
or even automatic parallelization.

\subsubsection{Code generation}
\index{Code generation}

The transformed intermediate language is translated into the output language, usually
the native machine language of the system. This involves resource and storage decisions,
such as deciding which variables to fit into registers and memory and the selection and
scheduling of appropriate machine instructions along with their associated addressing modes.

\section{Compilation vs. interpretation}
\index{Compilation vs. interpretation}

\subsection{Compilation}
\index{Compilation}

A compiler is a computer program (or set of programs) that transforms source code written
in a computer language (the source language) into another computer language (the target language,
often having a binary form known as object code). The most common reason for wanting to transform
source code is to create an executable program.

The name ``compiler'' is primarily used for programs that translate source code from a high-level
programming language to a lower level language (e.g., assembly language or machine code).
A program that translates from a low level language to a higher level one is a decompiler.
A program that translates between high-level languages is usually called a language translator,
source to source translator, or language converter. A language rewriter is usually a program
that translates the form of expressions without a change of language. While a typical compiler
outputs machine code from its final pass, there are several other types:

\subsubsection{Source-to-source compiler}
\index{Source-to-source compiler}

A source-to-source compiler is a type of compiler that takes a high level language
as its input and outputs a high-level language. For example, an automatic parallelizing
compiler will frequently take in a high-level language program as an input and then
transform the code and annotate it with parallel code annotations (e.g. OpenMP\index{OpenMP})
or language constructs (e.g. {\fortran}'s \api{DOALL} statements).

\subsubsection{Stage compiler}
\index{Stage compiler}

A ``stage compiler'' is a compiler that compiles to assembly language of a theoretical machine,
like most Prolog\index{Prolog} implementations that use what is known as the
``Warren Abstract Machine''\index{Warren Abstract Machine} (or WAM\index{WAM}),
designed by David Warren\index{David Warren} in 1983. Bytecode compilers for {\java}
(executed on a {\java} Virtual Machine (JVM\index{JVM})) and {\python} (executed on the CPython
virtual machine), and many more are also a subtype of this. In a sense, languages compiled
in this manner are evaluated in a hybrid compilation/interpretation mode, where the source code
is compiled into byte code, which is then interpreted at runtime.

\subsubsection{Dynamic compilation}
\index{Dynamic compilation}

Dynamic compilation is a process used by some programming language implementations
to gain performance during program execution. Although the technique originated
in the Sun's Self\index{Self} programming language, the best-known language that
uses this technique is {\java}. It allows optimizations to be made that can only
be known at runtime. Runtime environments using dynamic compilation typically
have programs run slowly for the first few minutes, and then after that, most of the
compilation and recompilation is done and it runs faster as execution goes.
Due to this initial performance lag, dynamic compilation is undesirable in certain cases.
In most implementations of dynamic compilation, some optimizations that could be done
at the initial compile time are delayed until further compilation at runtime,
causing further unnecessary slowdowns. Just-in-time compilation is a form of dynamic compilation.

In web development in this category, for example, JSP\index{JSP} \cite{jsp} pages and the associated tag libraries
e.g. within the Tomcat \cite{tomcat} container get first automatically compiled into
{\java} servlets (Java classes implemented the Servlet API \cite{servlets}) that then
dynamically compiled using the standard Java compiler into \file{.class} files.

\subsubsection{Just-in-time compiler}
\index{Just-in-time compiler}
\index{JIT}
\index{JIT!Java}
\index{JIT!.NET}
\index{Java!JIT}
\index{.NET!JIT}

Just-in-time compilation (JIT), also known as dynamic translation, is a technique
for improving the runtime performance of a computer program. JIT builds upon two earlier
ideas in run-time environments: bytecode compilation and dynamic compilation.
It converts code at runtime prior to executing it natively, for example bytecode
into native machine code. The performance improvement over interpreters originates
from caching the results of translating blocks of code, and not simply reevaluating
each line or operand each time it is met, such as in classical interpretation.
It also has advantages over statically compiling the code at development time,
as it can recompile the code if this is found to be advantageous, and may be able to
enforce security\index{security} guarantees. Thus JIT can combine some of
the advantages of interpretation and static (ahead-of-time) compilation.
Several modern runtime environments, such as Microsoft's .NET Framework
and most implementations of {\java}, rely on JIT compilation for high-speed code execution.

In a bytecode-compiled system, source code is translated to an intermediate
representation known as {\em bytecode} \index{bytecode}. Bytecode is not the machine code for any
particular computer, and may be portable among computer architectures. The bytecode may
then be interpreted by, or run on, a virtual machine. A just-in-time compiler can be used
as a way to speed up execution of bytecode. At the time the bytecode is run, the just-in-time
compiler will compile some or all of it to native machine code for better performance.
This can be done per-file, per-function or even on any arbitrary code fragment;
the code can be compiled when it is about to be executed (hence the name ``just-in-time'').

In contrast, a traditional interpreted virtual machine will simply interpret the bytecode,
generally with much lower performance. Some interpreters even interpret source code,
without the step of first compiling to bytecode, with even worse performance.
Statically compiled code or native code is compiled prior to deployment.
A dynamic compilation environment is one in which the compiler can be used during execution.
For instance, most Common LISP\index{Common LISP} systems have a compile function
which can compile new functions created at runtime. This provides many of the advantages
of JIT, but the programmer, rather than the runtime system, is in control of what parts
of the code are compiled. This can also compile dynamically generated code, which can,
in many scenarios, provide substantial performance advantages over statically compiled code,
as well as over most JIT systems.

A common goal of using JIT techniques is to reach or surpass the performance of static compilation,
while maintaining the advantages of bytecode interpretation: much of the ``heavy lifting''
of parsing the original source code and performing basic optimization is often handled at compile time,
prior to deployment: compilation from bytecode to machine code is much faster than compiling from source.
The deployed bytecode is portable, unlike native code. Compilers from bytecode to machine code are easier
to write, because the portable bytecode compiler has already done much of the analysis work.

JIT techniques generally offer far better performance than interpreters. In addition,
it can in some or many cases offer better performance than static compilation,
as many optimizations are only feasible at run-time, such as:

\begin{itemize}
	\item 
The compilation can be optimized to the targeted CPU and the
operating system model where the application runs.
	\item 
The system is able to collect statistics about how the program
is actually running in the environment it is in, and
it can rearrange and recompile for optimum performance.
	\item 
The system can do global code optimizations (e.g. inlining of
library functions) without losing the advantages of dynamic
linking and without the overheads inherent to static compilers and linkers.
	\item 
Although this is possible with statically compiled garbage collected languages,
a bytecode system can more easily rearrange memory for better cache utilization. 
\end{itemize}

\subsubsection{Ahead-of-time compile}
\index{Ahead-of-time compile}

Ahead-of-time compilation (AOT) refers to the act of compiling an intermediate language,
such as {\java} bytecode or .NET Common Intermediate Language (CIL\cite{CIL}),
into a system-dependent binary. Most languages that can be compiled to an intermediate
language (such as bytecode) take advantage of just-in-time compilation.
JIT compiles intermediate code into binary code for a native run while the intermediate
code is executing, which may decrease an application's performance. Ahead-of-time
compilation eliminates the need for this step by performing the compilation before
execution rather than during execution.

\subsection{Interpretation}
\index{Interpretation}

An interpreted language is a programming language whose programs are not directly executed
by the host cpu but rather executed (or said to be interpreted) by a software program known
as an interpreter. The source code of the program is often translated to a form that is more
convenient to interpret, which may be some form of machine language for a virtual machine
(such as {\java}'s bytecode). Theoretically, any language may be compiled or interpreted,
so this designation is applied purely because of common implementation practice
and not some underlying property of a language. 

Many languages have been implemented using both compilers and interpreters,
including {\lisp}, Pascal\index{Pascal}, {\C}, BASIC, and {\python}.
While {\java} is translated to a form that is intended to be interpreted,
just-in-time compilation is often used to generate machine code at run time.
The Microsoft's .NET languages compile to CIL (Microsoft's Common Intermediate Language)
which is often then compiled into native machine code; however, there is a virtual
machine capable of interpreting CIL. Many {\lisp} implementations can freely mix
interpreted and compiled code. These implementations also use a compiler
that can translate arbitrary source code at runtime to machine code.

In the early days of computing, language design was heavily influenced by the decision
to use compilation or interpretation as a mode of execution. For example,
some compiled languages require that programs must explicitly state the data-type
of a variable at the time it is declared or first used while some interpreted languages
take advantage of the dynamic aspects of interpretation to make such declarations unnecessary.
For example, Smalltalk 80\index{Smalltalk!80}, which was designed to be interpreted
at run-time, allows generic Objects to dynamically interact with each other.

Initially, interpreted languages were compiled line-by-line; that is, each line was compiled
as it was about to be executed, and if a loop or subroutine caused certain lines to be
executed multiple times, they would be recompiled every time. This has become much less common.
Most so-called interpreted languages use an intermediate representation,
which combines both compilation and interpretation. In this case, a compiler
may output some form of bytecode or threaded code, which is then executed by a bytecode
interpreter. Examples include {\python}, {\java}, and Ruby\index{Ruby}. The intermediate
representation can be compiled once and for all (as in {\java}), each time before execution
(as in {\perl} or Ruby), or each time a change in the source is detected
before execution (as in {\python}).

Interpreting a language gives implementations some additional flexibility
over compiled implementations. Features that are 
often easier to implement in interpreters than in compilers include (but are not limited to):

\begin{itemize}
	\item 
platform independence ({\java}'s bytecode, for example)
	\item 
reflection and reflective usage of the evaluator
	\item 
dynamic typing
	\item 
smaller executable program size (since implementations have flexibility to choose the instruction code)
	\item 
dynamic scoping
\end{itemize}

The main disadvantage of interpreting is a much slower speed
of program execution compared to direct machine code execution
on the host CPU. A technique used to improve performance
is just-in-time compilation which converts frequently executed
sequences of interpreted instruction to host machine code.

\subsection{Subreferences}

\begin{enumerate}
	\item 
\url{http://en.wikipedia.org/wiki/Semantic_analysis_(computer_science)}

	\item 
\url{http://en.wikipedia.org/wiki/Lexical_analysis}

	\item 
\url{http://en.wikipedia.org/wiki/Parsing}

	\item 
\url{http://en.wikipedia.org/wiki/Interpreted_language}

	\item 
\url{http://en.wikipedia.org/wiki/Dynamic_compilation}

	\item 
\url{http://en.wikipedia.org/wiki/Just-in-time_compilation}

	\item 
\url{http://en.wikipedia.org/wiki/Compiler}
\end{enumerate}

\section{Type System}
\index{Type System}

A type system is a framework for classifying programming languages'
{\em phrases} according to the kinds of values they compute. A type system associates
a type with each computed value. By examining the flow of these values, a type system
attempts to prove that no type errors can occur in a given program. The type system
in question determines what constitutes a type error, but a type system generally seeks
to guarantee that operations expecting a certain kind of value are not used with
values for which that operation makes no sense.

Technically, assigning data types (i.e. {\em typing}) gives meaning to collections of bits
in the computer's memory. Types usually have associations either with values in memory
or with objects such as variables. Because any value simply consists of a sequence
of bits in a computer, hardware makes no intrinsic distinction even between memory addresses,
instruction code, characters, integers and floating-point numbers, being unable to discriminate
between them based on bit pattern alone. Associating a sequence of bits and a type
informs programs and programmers how that sequence of bits should be understood,
i.e. it gives semantic information to the values manipulated by a computer program.
Major functions provided by type systems include:

\begin{itemize}
	\item 
{\bf Safety} -- Use of types may allow a compiler to detect meaningless or probably invalid code.
For example, we can identify an expression:

\begin{verbatim}
3 / ``Hello, World''
\end{verbatim}

as invalid (depending on the language) because the rules of arithmetic do not specify
how to divide an integer by a string. As discussed below, strong typing offers more safety,
but generally does not guarantee complete safety.

	\item 
{\bf Optimization} -- Static type-checking may provide useful compile-time information.
For example, if a type requires that a value must align in memory at a multiple of 4 bytes,
the compiler may be able to use more efficient machine instructions.

	\item 
{\bf Documentation} -- In more expressive type systems, types can serve as a form of documentation,
since they can illustrate the intent of the programmer. For instance, timestamps may be represented
as integers, but if a programmer declares a function as returning a timestamp type rather than merely
an integer type, this documents part of the meaning of the function.

	\item 
{\bf Abstraction} -- Types allow programmers to think about programs at a higher level
than the bit or byte, not bothering with low-level implementation. For example, programmers
can think of a string as a collection of character values instead of as a mere array of bytes.
Or, types can allow programmers to express the interface between two subsystems.
This helps localize the definitions required for interoperability of the subsystems and prevents
inconsistencies when those subsystems communicate.
\end{itemize}

\subsection{Type checking}
\index{Type checking}

The process of verifying and enforcing the constraints of types -- type checking --
may occur either at compile-time (a static check) or run-time (a dynamic check).
If a language specification requires its typing rules strongly (i.e. more or less
allowing only those automatic type conversions which do not lose information),
one can refer to the process as {\em strongly typed}, if not, as {\em weakly typed}. 

\subsubsection{Static typing}
\index{Static typing}

A programming language is said to use static typing when type checking is
performed during compile-time as opposed to run-time. Statically typed languages
include Ada, {\C}, {\cpp}, {\csharp}, F\#, {\java}, {\fortran}, {\haskell}, ML,
Pascal, {\perl}, Objective-C and Scala.
Static typing is a limited form of program verification. It allows many type errors
to be caught early in the development cycle. Static type checkers evaluate
only the type information that can be determined at compile time, but are able to verify
that the checked conditions hold for all possible executions of the program,
which eliminates the need to repeat type checks every time the program is executed.
Program execution may also be made more efficient (i.e. faster or taking reduced memory)
by omitting runtime type checks and enabling other optimizations.

Because they evaluate type information during compilation, and therefore lack type
information that is only available at run-time, static type checkers are conservative.
They will reject some programs that may be well-behaved at run-time,
but that cannot be statically determined to be well-typed. For example,
even if an expression \texttt{<complex test>} always evaluates to true at run-time,
a program containing the code

\begin{verbatim}
    if <complex test> then 42 else <type error>
\end{verbatim}

\noindent
will be rejected as ill-typed, because a static analysis cannot determine that
the \api{else} branch won't be taken. The conservative behaviour of static type checkers
is advantageous when \texttt{<complex test>} evaluates to \api{false} infrequently:
a static type checker can detect type errors in rarely used code paths. 

Note that the most widely used statically typed languages are not formally type safe.
They have ``loopholes'' in the programming language specification enabling programmers
to write code that circumvents the verification performed by a static type checker.
For example, most {\C}-style languages have type coercion, and {\haskell}
has such features as \api{unsafePerformIO}: such operations may be unsafe at runtime,
in that they can cause unwanted behaviour due to incorrect typing of values when the program runs.

\subsubsection{Dynamic typing}
\index{Dynamic typing}

A programming language is said to be dynamically typed when the majority of its type checking is performed at run-time as opposed to at compile-time. In dynamic typing, values have types but variables do not; that is, a variable can refer to a value of any type. Dynamically typed languages include Erlang, Groovy, JavaScript, {\lisp}, Objective-C, {\perl} (with respect to user-defined types but not built-in types), {\php}, Prolog, {\python}, Ruby and Smalltalk. Compared to static typing, dynamic typing can be more flexible (e.g. by allowing programs to generate types and functionality based on run-time data), though at the expense of fewer a priori typing guarantees. This is because a dynamically typed language accepts and attempts to execute some programs which may be ruled as invalid by a static type checker. 

Dynamic typing may result in runtime type errors—that is, at runtime, a value may have an unexpected type, and an operation nonsensical for that type is applied. This operation may occur long after the place where the programming mistake was made—that is, the place where the wrong type of data passed into a place it should not have. This may make the bug difficult to locate.

Dynamically typed language systems, compared to their statically typed cousins, make fewer ``compile-time'' checks on the source code (but will check, for example, that the program is syntactically correct). Run-time checks can potentially be more sophisticated, since they can use dynamic information as well as any information that was present during compilation. On the other hand, runtime checks only assert that conditions hold in a particular execution of the program, and these checks are repeated for every execution of the program.

Development in dynamically typed languages is often supported by programming practices such as unit testing. Testing is a key practice in professional software development, and is particularly important in dynamically typed languages. In practice, the testing done to ensure correct program operation can detect a much wider range of errors than static type-checking, but conversely cannot search as comprehensively for the errors that both testing and static type checking are able to detect.

\subsubsection{Duck typing}
\label{sect:duck-typing}
\index{Duck typing}

Duck typing is a style of dynamic typing in which an object's current
set of methods and properties determines the valid semantics,
rather than its inheritance from a particular class or implementation
of a specific interface. The name of the concept refers to the duck test,
attributed to American writer James Whitcomb Riley, which may be phrased as follows:

\begin{quote}
    ``when I see a bird that walks like a duck and swims like a duck and quacks
    like a duck, I call that bird a duck.''
\end{quote}

In duck typing, one is concerned with just those aspects of an object that are used,
rather than with the type of the object itself. For example, in a non-duck-typed language,
one can create a function that takes an object of type \api{Duck} and calls that object's
\api{walk()} and \api{quack()} methods. In a duck-typed language, the equivalent
function would take an object of any type and call that object's \api{walk()} and \api{quack()} methods.
If the object does not have the methods that are called then the function signals a run-time error.
It is this action of any object having the correct \api{walk()} and \api{quack()} methods being accepted
by the function that evokes the quotation and hence the name of this form of typing.

Duck typing is aided by habitually not testing for the type of arguments
in method and function bodies, relying on documentation, clear code,
and unit testing to ensure correct use. Users of statically typed languages
new to dynamically typed languages are usually tempted to add such static
(before run-time) type checks, defeating the benefits and flexibility of duck typing,
and constraining the language's dynamism.

\subsubsection{Structural type system}
\index{Structural type system}

A structural type system (or property-based type system)
is a major class of type systems, in which type compatibility
and equivalence are determined by the type's structure,
and not through explicit declarations. Structural systems are used to
determine if types are equivalent, as well as if a type
is a subtype of another. It contrasts with nominative systems (see \xs{sect:nominative-type-system}),
where comparisons are based on explicit declarations or the names
of the types, and duck typing (see \xs{sect:duck-typing}), in which only the part 
f the structure accessed at runtime is checked for compatibility.

In structural typing, two objects or terms are considered
to have compatible types if the types have identical structure.
Depending on the semantics of the language, this generally means
that for each feature within a type, there must be a corresponding
and identical feature in the other type. Some languages may differ
on the details (such as whether the features must match in name).

ML and Objective Caml are examples of structurally-typed languages.
{\cpp} template functions exhibit structural typing on type arguments. 

In languages, which support subtype polymorphism, a similar dichotomy
can be formed based on how the subtype relationship is defined.
{\em One type is a subtype of another if and only if it contains all the features
of the base type (or subtypes thereof); the subtype may contain additional
features (such as members not present in the base type, or stronger invariants).}

A pitfall of structural typing versus nominative typing is that two separately
defined types intended for different purposes, each consisting of a pair of numbers,
could be considered the same type by the type system, simply because they happen
to have identical structure. One way this can be avoided is by creating one algebraic
data type for one use of the pair and another algebraic data type for the other use.

\subsubsection{Nominative type system}
\label{sect:nominative-type-system}
\index{Nominative type system}

A nominal or nominative type system (or name-based type system) is a major class of type system, in which compatibility and equivalence of data types is determined by explicit declarations and/or the name of the types. Nominative systems are used to determine if types are equivalent, as well as if a type is a subtype of another. It contrasts with structural systems, where comparisons are based on the structure of the types in question and do not require explicit declarations.

Nominal typing means that two variables are type-compatible if and only if their declarations name the same type. For example, in C, two struct types with different names are never considered compatible, even if they have identical field declarations. However, C also allows a typedef declaration, which introduces an alias for an existing type. These are merely syntactical and do not differentiate the type from its alias for the purpose of type checking. 

In a similar fashion, nominal subtyping means that one type is a subtype of another if and only if it is explicitly declared to be so in its definition. Nominally-typed languages typically enforce the requirement that declared subtypes be structurally compatible (though Eiffel allows non-compatible subtypes to be declared). However, subtypes which are structurally compatible ``by accident'', but not declared as subtypes, are not considered to be subtypes. C, {\cpp}, {\csharp} and {\java} all primarily use both nominal typing and nominal subtyping.

Some nominally-subtyped languages, such as {\java} and {\csharp}, allow classes to be declared final (or sealed in {\csharp} terminology), indicating that no further subtyping is permitted.

Nominal typing is useful at preventing accidental type equivalence, and is considered to have better type-safety than structural typing. The cost is a reduced flexibility, as, for example, nominal typing does not allow new super-types to be created without modification of the existing subtypes. 

\begin{enumerate}
	\item 
\url{http://en.wikipedia.org/wiki/Type_system}

	\item 
\url{http://en.wikipedia.org/wiki/Duck_typing}

	\item 
\url{http://en.wikipedia.org/wiki/Dynamic_typing}

	\item 
\url{http://en.wikipedia.org/wiki/Type_safety}

	\item 
\url{http://en.wikipedia.org/wiki/Structural_type_system}

	\item 
\url{http://en.wikipedia.org/wiki/Nominative_type_system}
\end{enumerate}

\section{Memory management}
\index{Memory management}

Memory management is the act of managing computer memory. In its simpler forms, this involves providing ways to allocate portions of memory to programs at their request, and freeing it for reuse when no longer needed. The management of main memory is critical to the computer system.

Garbage collection is the automated allocation, and deallocation of computer memory resources for a program. This is generally implemented at the programming language level and is in opposition to manual memory management, the explicit allocation and deallocation of computer memory resources.

\subsection{Garbage collection}
\index{Garbage collection}

Garbage collection is a form of automatic memory management. It is a special case of resource management, in which the limited resource being managed is memory. The garbage collector attempts to reclaim garbage, or memory occupied by objects that are no longer in use by the program. Garbage collection was invented by John McCarthy around 1959 to solve problems in Lisp.

Garbage collection is often portrayed as the opposite of manual memory management, which requires the programmer to specify which objects to deallocate and return to the memory system. However, many systems use a combination of the two approaches, and other techniques such as stack allocation and region inference using syntactical program blocks can carve off parts of the problem. 

Garbage collection does not traditionally manage limited resources other than memory that typical programs use, such as network sockets, database handles, user interaction windows, and file and device descriptors. Methods used to manage such resources, particularly destructors, may suffice as well to manage memory, leaving no need for garbage collection. Some garbage collection systems allow such other resources to be associated with a region of memory that, when collected, causes the other resource to be reclaimed; this is called finalization. Finalization may introduce complications limiting its usability, such as intolerable latency between disuse and reclaim of especially limited resources.

A finalizer is a special method that is executed when an object is garbage collected. It is similar in function to a destructor. In less technical terms, a finalizer is a piece of code that ensures that certain necessary actions are taken when an acquired resource (such as a file or access to a hardware device) is no longer being used. This could be closing the file or signaling to the operating system that the hardware device is no longer needed. However, as noted below, finalizers are not the preferred way to accomplish this and for the most part serve as a fail-safe. Unlike destructors, finalizers are not deterministic. A destructor is run when the program explicitly frees an object. A finalizer, by contrast, is executed when the internal garbage collection system frees the object. Programming languages which use finalizers include {\java} and {\csharp}. In {\csharp}, and a few others which support finalizers, the syntax for declaring a finalizer mimics that of destructors in {\cpp}.

Due to the lack of programmer control over their execution, it is usually recommended to avoid finalizers for any but the most trivial operations. In particular, operations often performed in destructors are not usually appropriate for finalizers. For example, destructors are often used to free expensive resources such as file or network handles. If placed in a finalizer, the resources may remain in use for long periods of time after the program is finished with them. Instead, most languages encourage the dispose pattern whereby the object has a method to clean up the object's resources, leaving the finalizer as a fail-safe in the case where the dispose method doesn't get called. The {\csharp} language supports the dispose pattern explicitly, via the \api{IDisposable} interface.

The basic principles of garbage collection are:

\begin{enumerate}
	\item 
 Find data objects in a program that cannot be accessed in the future
	\item 
 Reclaim the resources used by those objects
\end{enumerate}

By making manual memory deallocation unnecessary (and often forbidding it), garbage collection frees the programmer from having to worry about releasing objects that are no longer needed, which can otherwise consume a significant amount of design effort. It also aids programmers in their efforts to make programs more stable, because it prevents several classes of runtime errors. For example, it prevents dangling pointer errors, where a reference to a deallocated object is used. The pointer still points to the location in memory where the object or data was, even though the object or data has since been deleted and the memory may now be used for other purposes. This can, and often does, lead to storage violation errors that are extremely difficult to resolve.

Many computer languages require garbage collection, either as part of the language specification (e.g., {\java}, {\csharp}, and most scripting languages) or effectively for practical implementation (e.g., formal languages like lambda calculus); these are said to be garbage collected languages. Other languages were designed for use with manual memory management, but have garbage collected implementations available (e.g., {\C}, {\cpp}). Some languages, like Ada, Modula-3, and {\cpp} allow both garbage collection and manual memory management to co-exist in the same application by using separate heaps for collected and manually managed objects; others, like D, are garbage collected but allow the user to manually delete objects and also entirely disable garbage collection when speed is required. 

\paragraph{Benefits}
\index{Garbage collection!Benefits}

Garbage collection frees the programmer from manually dealing with memory allocation and deallocation. As a result, certain categories of bugs are eliminated or substantially reduced:

\begin{itemize}
	\item 
 Dangling pointer bugs, which occur when a piece of memory is freed while there are still pointers to it, and one of those pointers is then used.
	\item 
 Double free bugs, which occur when the program attempts to free a region of memory that is already free.
	\item 
 Certain kinds of memory leaks, in which a program fails to free memory occupied by objects that will not be used again, leading, over time, to memory exhaustion.
\end{itemize}

\paragraph{Disadvantages}
\index{Garbage collection!Disadvantages}

Typically, garbage collection has certain disadvantages.

\begin{itemize}
	\item 
 Garbage collection is a process that consumes limited computing resources in deciding what memory is to be freed and when, reconstructing facts that may have been known to the programmer. The penalty for the convenience of not annotating memory usage manually in the code is overhead leading, potentially, to decreased performance.

	\item 
 The point when the garbage is actually collected can be unpredictable, resulting in delays scattered throughout a session. Unpredictable delays can be unacceptable in real-time environments such as device drivers, or in transaction processing. Recursive algorithms that take advantage of automatic storage management often delay automatic release of stack objects until after the final call has completed, causing increased memory requirements.

	\item 
 Memory may leak despite the presence of a garbage collector if references to unused objects are not themselves manually disposed of. Researchers draw a distinction between ``physical'' and ``logical'' memory leaks. In a physical memory leak, the last pointer to a region of allocated memory is removed, but the memory is not freed. In a logical memory leak, a region of memory is still referenced by a pointer, but is never actually used.[4] Garbage collectors generally can do nothing about logical memory leaks.

	\item 
 Perhaps the most significant problem is that programs that rely on garbage collectors often exhibit poor locality (interacting badly with cache and virtual memory systems), occupying more address space than the program actually uses at any one time, and touching otherwise idle pages. These may combine in a phenomenon called thrashing, in which a program spends more time copying data between various grades of storage than performing useful work. They may make it impossible for a programmer to reason about the performance effects of design choices, making performance tuning difficult. They can lead garbage-collecting programs to interfere with other programs competing for resources.

	\item 
 The execution of a program using a garbage collector is not deterministic. An object which becomes eligible for garbage collection will usually be cleaned up eventually, but there is no guarantee when (or even if) that will happen. Most run-time systems using garbage collectors require manual deallocation of limited non-memory resources (using finalizers), as an automatic deallocation during the garbage collection phase may run too late or in the wrong circumstances. Also, the performance impact caused by the garbage collector is seemingly random and hard to predict, leading to program execution that non-deterministic. 
\end{itemize}

Generally speaking, higher-level programming languages are more likely to have garbage collection as a standard feature. In languages that do not have built in garbage collection, it can often be added through a library, as with the Boehm garbage collector for {\C} and {\cpp}. This approach is not without drawbacks, such as changing object creation and destruction mechanisms.

Most functional programming languages, such as ML, {\haskell}, and APL, have garbage collection built in. Lisp, which introduced functional programming, is especially notable for introducing this mechanism.

Other dynamic languages, such as Ruby (but not {\perl}, or {\php}, which use reference counting), also tend to use garbage collection. Object-oriented programming languages such as Smalltalk and {\java} usually provide integrated garbage collection. A notable exception is {\cpp} which, uniquely, relies on destructors \cite{aspects-mem-management-java-cpp}.

\begin{enumerate}
	\item 
\url{http://en.wikipedia.org/wiki/Garbage_collection_%28computer_science%29}

	\item 
\url{http://en.wikipedia.org/wiki/Finalizer}
\end{enumerate}

\subsection{Manual memory management}

Manual memory management refers to the usage of manual instructions by the programmer to identify and deallocate unused objects, or garbage. Up until the mid 1990s, the majority of programming languages used in industry supported manual memory management. Today, however, languages with garbage collection are becoming increasingly popular; the main manually-managed languages still in widespread use today are {\C} and {\cpp}.

Most programming languages use manual techniques to determine when to allocate a new object from the free store. {\C} uses the \api{malloc()} function; {\cpp} and {\java} use the \api{new} operator; determination of when an object ought to be created is trivial and unproblematic. The fundamental issue is determination of when an object is no longer needed (ie. is garbage), and arranging for its underlying storage to be returned to the free store so that it may be re-used to satisfy future memory requests. In manual memory allocation, this is also specified manually by the programmer; via functions such as \api{free()} in {\C}, or the delete operator in {\cpp}. Manual memory management is known to enable several major classes of bugs into a program, when used incorrectly:

\begin{itemize}
	\item 
 When an unused object is never released back to the free store, this is known as a memory leak. In some cases, memory leaks may be tolerable, such as a program which ``leaks'' a bounded amount of memory over its lifetime, or a short-running program which relies on an operating system to deallocate its resources when it terminates. However, in many cases memory leaks occur in long-running programs, and in such cases an unbounded amount of memory is leaked. When this occurs, the size of the available free store continues to decrease over time; when it finally is exhausted the program then crashes.

	\item 
 When an object is deleted more than once, or when the programmer attempts to release a pointer to an object not allocated from the free store, catastrophic failure of the dynamic memory management system can result. The result of such actions can include heap corruption, premature destruction of a different (and newly-created) object which happens to occupy the same location in memory as the multiply-deleted object, and other forms of undefined behavior.

	\item 
 Pointers to deleted objects become wild pointers if used post-deletion; attempting to use such pointers can result in difficult-to-diagnose bugs.

Languages which exclusively use garbage collection are known to avoid the last two classes of defects. Memory leaks can still occur (and bounded leaks frequently occur with generational or conservative garbage collection), but are generally less severe than memory leaks in manual memory allocation schemes.

Manual memory management has one correctness advantage, which comes into play when objects own scarce system resources (like graphics resources, file handles, or database connections) which must be relinquished when an object is destroyed. Languages with manual management, via the use of destructors, can arrange for such actions to occur at the precise time of object destruction. In {\cpp}, this ability is put to further use to automate memory deallocation within an otherwise-manual framework, use of the auto ptr template in the language's standard library to perform memory management is a common paradigm. 

Many advocates of manual memory management argue that it affords superior performance when compared to automatic techniques such as garbage collection. Manual allocation does not suffer from the long ``pause'' times often associated with garbage collection (although modern garbage collectors have collection cycles which are often not noticeable), and manual allocation frequently has superior locality of reference. This is especially an issue in real time systems, where unbounded collection cycles are generally unacceptable. Manual allocation is also known to be more appropriate for systems where memory is a scarce resource. 

On the other hand, manual management has documented performance disadvantages:

	\item 
 Calls to delete and such incur an overhead each time they are made, this overhead can be amortized in garbage collection cycles. This is especially true of multithreaded applications, where delete calls must be synchronized.

	\item 
 The allocation routine may be more complicated, and slower. Some garbage collection schemes, such as those with heap compaction, can maintain the free store as a simple array of memory (as opposed to the complicated implementations required by manual management schemes).
\end{itemize}

\begin{enumerate}
	\item 
\url{http://en.wikipedia.org/wiki/Manual_memory_management}
\end{enumerate}

\chapter{Programming Languages Popularity}
\index{Popularity}

Since the beginning of the computer, thousands of programming languages have been designed.
However, a relatively few proportion of these were actually put to practical use in the industry.
Some languages have been developed and/or adopted by the academic community for teaching
purposes but did not really come into practical use either. Due to their adoption in University
computer science programs, such languages can boast great popularity, but still failed
to have become popular for extensive industrial software development. Many factors or
indicators can be used to measure the popularity of programming languages. Here we discuss
various surveys that have been made to measure the popularity of programming languages.

\section{langpop.com}

The methodology and results presented here are based on: \url{http://www.langpop.com/}.
It uses various web sites as a data source to extract indicators of programming languages'
popularity rankings.

\subsection{Yahoo Search}

Arguably, the popularity of a programming language can be measured by the number of web pages
discussing it. Yahoo provides an API for doing customized searches on its web pages indexing database.
Searching for various programming languages' names, \url{http://www.langpop.com/} arrives to the conclusion
that the programming languages most referred to on the web are: {\cpp}, {\C}, {\java}, and {\php},
as shown on the extracted figure:

\includegraphics[width=\textwidth]{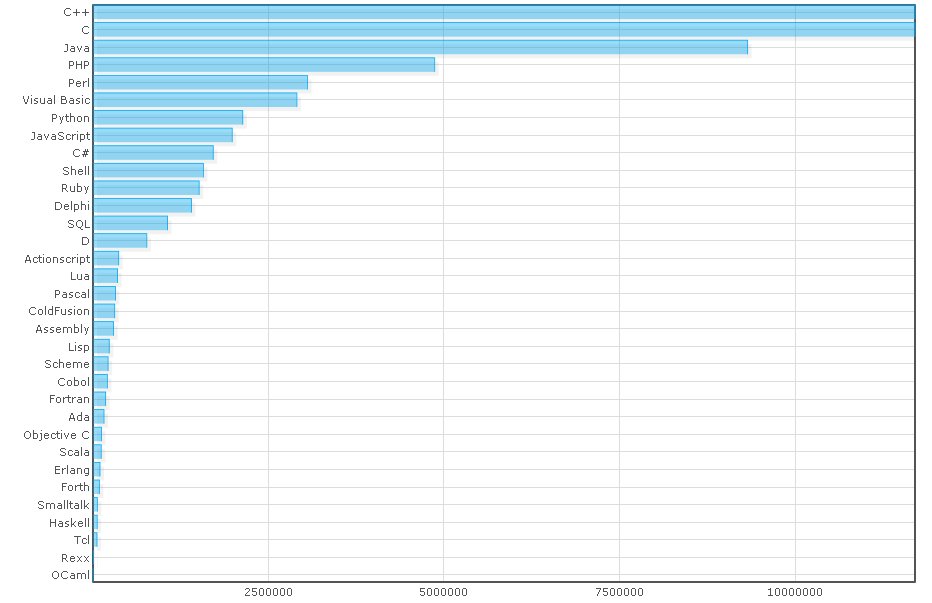}

\subsection{Job Postings}

Since industry demand is probably a more relevant indicator of a language's popularity,
one might search a job offerings database and search for specific programming languages
being asked for in job offerings. Based on a search of \url{craigslist.org}, \url{http://www.langpop.com/}
arrives to the conclusion that {\php}, SQL, {\C}, {\cpp}, {\java}, JavaScript, and {\csharp}
are the most in-demand languages, as shown on the extracted figure:

\includegraphics[width=\textwidth]{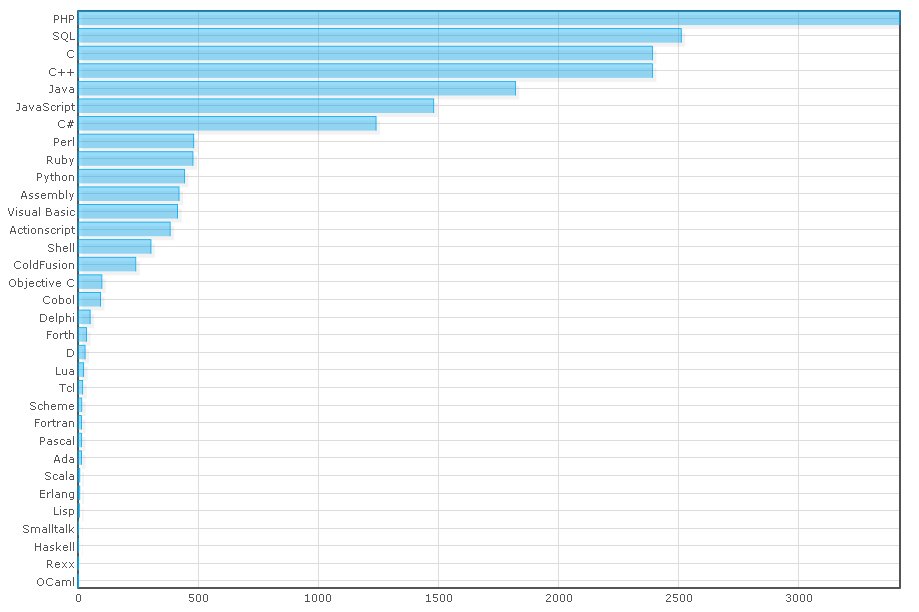}

\subsection{Books}

Arguably, the more a programming language is popular, the more books are being sold
explaining how to use it. In \url{http://www.langpop.com/}, a search is made to a bookstore
database to figure out what languages are more popular as books' topics.
It arrives to the conclusion that {\java}, {\cpp}, Visual Basic, {\csharp}, and {\C}
are the most popular from this regard, as shown on the extracted figure:

\includegraphics[width=\textwidth]{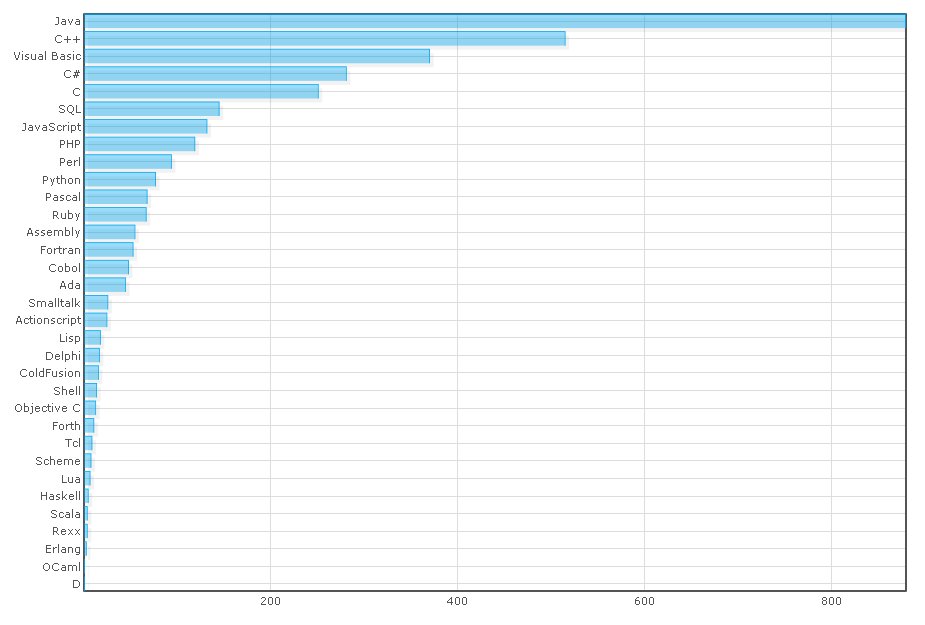}

\subsection{Open source code repositories}

The open source code development community provides a public a free source of information
that can be used to measure the popularity of programming languages effectively used by developers.
In \url{http://www.langpop.com/}, a search is made on various open source code repositories
and the programming language used in each open source project is extracted.

\subsubsection{freshmeat.org}

While searching the \url{freshmeat.net} open source community web site,
it showed that {\C}, {\java}, {\cpp}, {\php}, Perl, and Python are the most popular, as shown in the extracted figure:

\includegraphics[width=\textwidth]{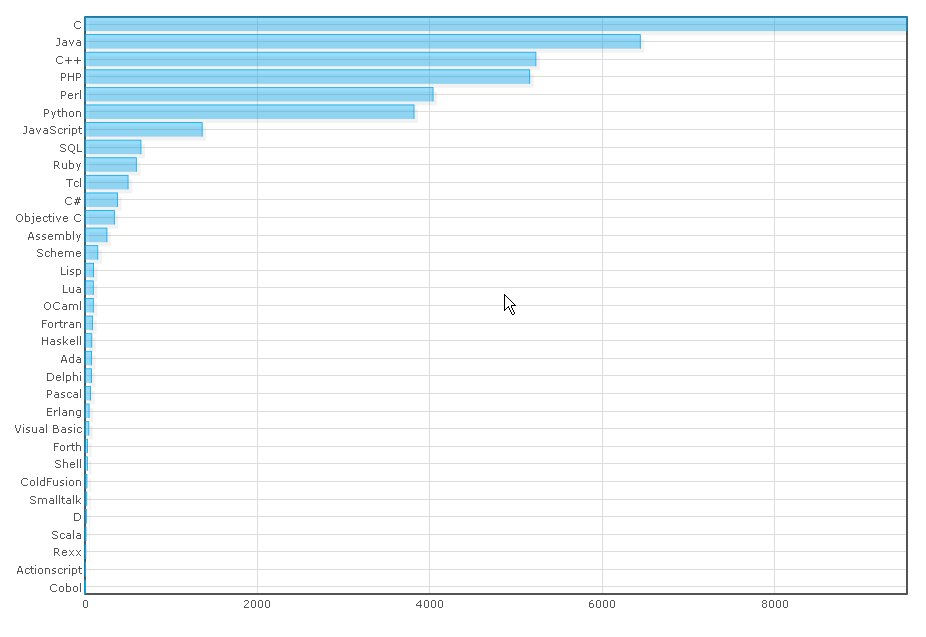}

\subsubsection{Google code search}

Doing a similar search on the Google code web site (\url{http://www.google.com/codesearch}),
it was extracted that (as for the results for freshmeat.org) C, Java, C++, PHP, Perl, and Python are the most popular languages.

\subsubsection{Ohloh}

Doing yet another similar search on the Ohloh web site (www.ohloh.net) \cite{ohloh}, it was extracted that, quite differently from the two previous results, Java, C, C++, JavaScript, Shell, Python, and PHP are the most popular languages, as shown in the figure:

\includegraphics[width=\textwidth]{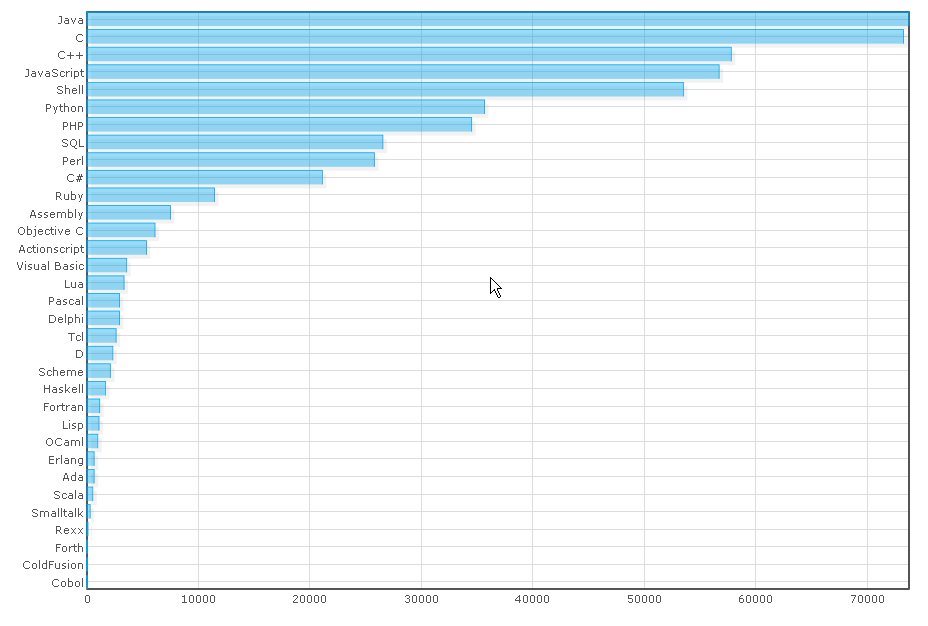}

\subsection{Discussion sites}

Another relevant source of information to measure the relative popularity of programming languages is public discussion sites, where programmers discuss technical issues among themselves. In \url{http://www.langpop.com/}, the following discussion sites were searched for: \url{lambda-the-ultimate.org}, \url{slashdot.org}, \url{programming.reddit.com}, and \url{freenode.net}. Overall, the results of these searches revealed that Java, C, C++, C\#, PHP and Python are the languages that the programmers communities are the most discussing among themselves, as shown in the extracted figure:

\includegraphics[width=\textwidth]{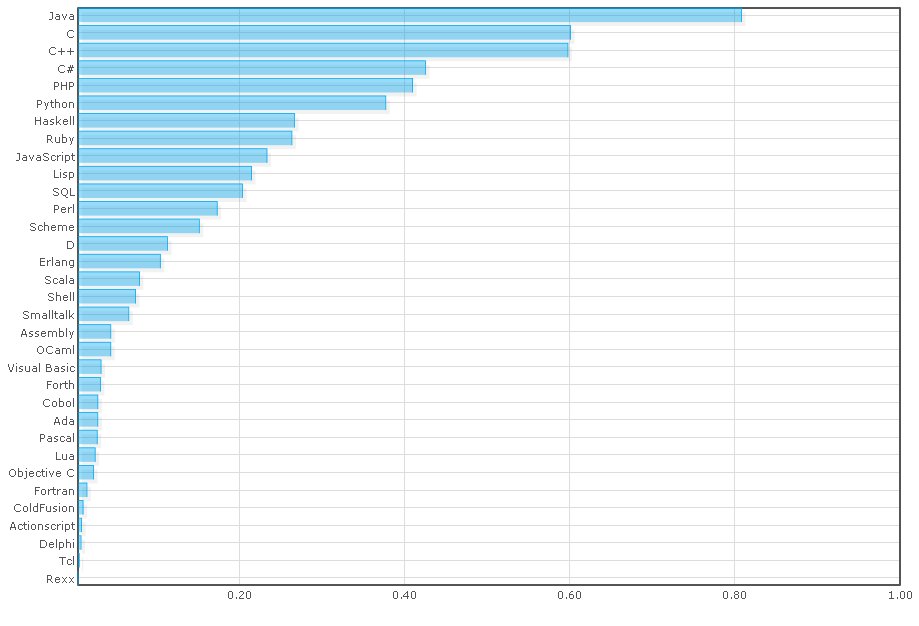}. 

Notably in the search on public discussion sites, \url{lambda-the-ultimate.org} is a site that is very popular particularly with the academia circles. Searches on this site has revealed a different taste in academia, showing that languages like Haskell, Lisp, Scheme, Python, Erlang, Ruby, Scala, Smalltalk, and Ada are popular discussion topics in academic circles, as shown in the extracted figure:

\includegraphics[width=\textwidth]{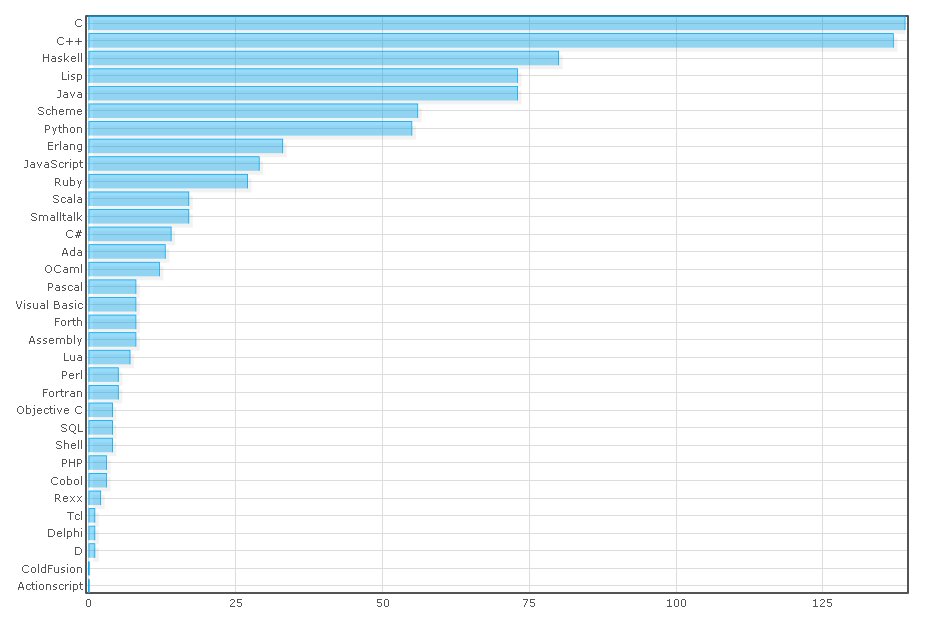}.

\section{www.tiobe.com}

The results presented here are based on a monthly survey done by Tiobe Software (\url{www.tiobe.com}). As stated on the ``Tiobe index'' web page updated monthly, the TIOBE Programming Community index gives an indication of the popularity of programming languages. The index is updated once a month. The ratings are based on the number of skilled engineers world-wide, courses and third party vendors. The popular search engines Google, MSN, Yahoo!, Wikipedia and YouTube are used to calculate the ratings. Observe that the TIOBE index is not about the best programming language or the language in which most lines of code have been written. The data presented here is as of March 2010. 

\subsection{Top 20 programming languages (March 2010)}

According to the TIOBE index (March 2010), the top 20 most popular programming languages are depicted as in the extracted figure:

\includegraphics[width=\textwidth]{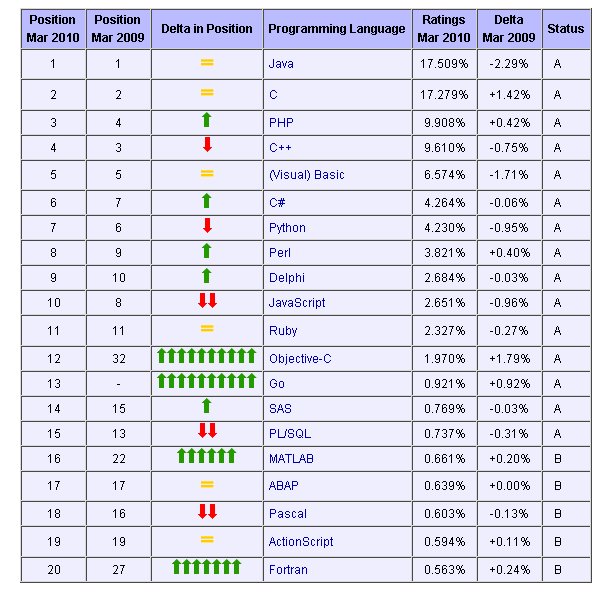}.

\subsection{Long-term trend}

As the TIOBE index is calculated monthly, and has been there for a long time, it is very interesting to see long-term trends in the history of programming languages popularity, as depicted in the extracted figure:

\includegraphics[width=\textwidth]{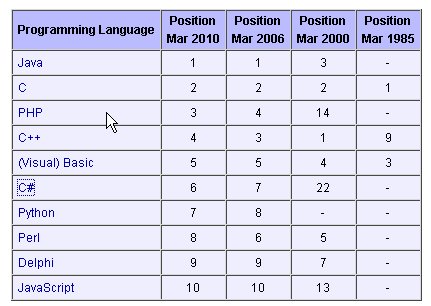}. 

\subsection{Categories of programming languages}

The TIOBE index also enables the categorization of popular programming languages. The results are as depicted in the extracted figure:

\includegraphics[width=\textwidth]{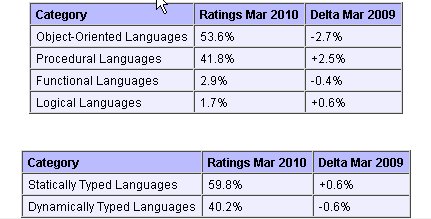}.


\chapter{Programming Languages Performance Ranking}
\index{Performance}

In order to perform a relative evaluation of programming languages
other than based on their popularity, the execution performance of
programs can be exercised and compared. This requires the design
of algorithmically comparable solutions in various languages and
executing these programs in the same execution environment.

Many such empirical studies have been made to compare programming languages.
Notably, the Computer Languages Benchmark Game [1] (\url{http://shootout.alioth.debian.org/})
provides a dynamic web site demonstrating extensive empirical results enabling
the comparison of various programming languages using a wide variety of benchmarking programs.

Among the most revealing comparisons that can be found on this site is the following:
[2] (\url{http://shootout.alioth.debian.org/u32q/which-language-is-best.php}).
It enables the comparison of programming languages according to performance criteria such as:
execution time, memory consumption, and source code size. 

\section{Execution time}

\includegraphics{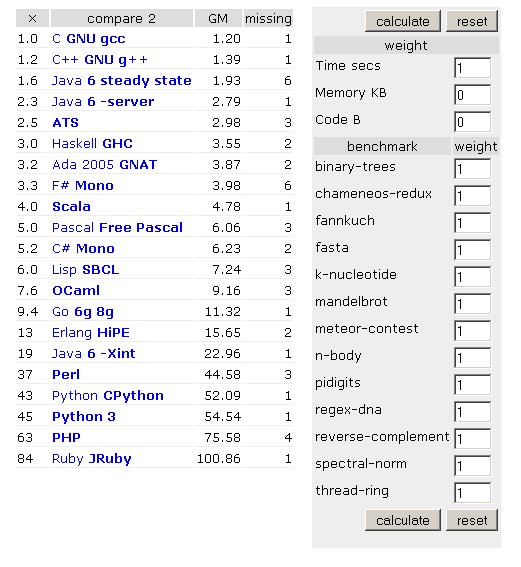}

\section{Memory consumption}

\includegraphics{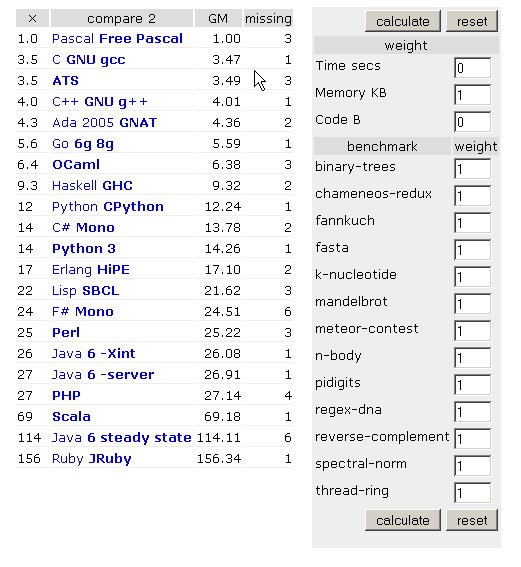}

\section{Source code size}

\includegraphics{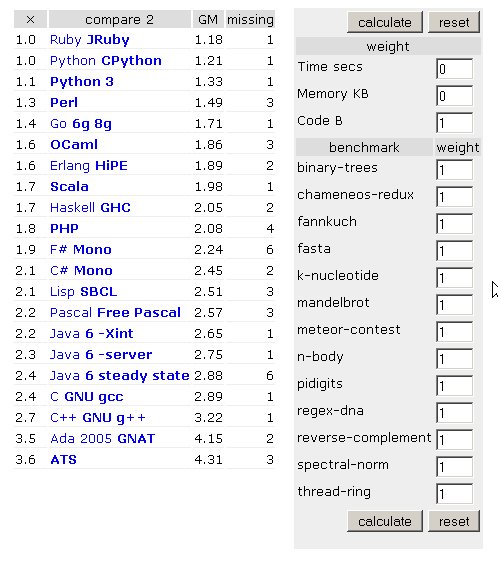}

\section{Overall results}

\includegraphics{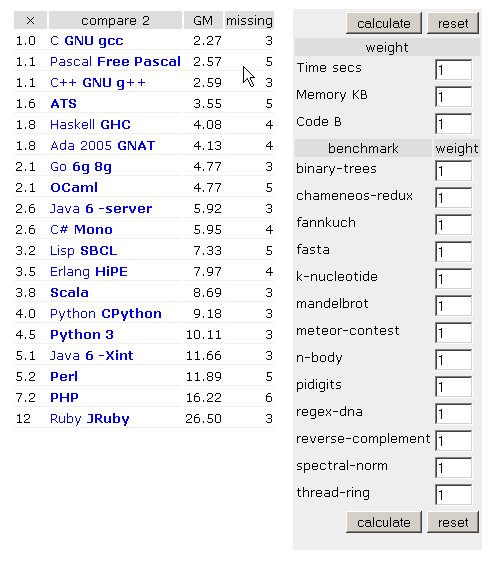}


\bibliographystyle{plain}
\bibliography{../../common/bibliography}

\begin{thebibliography}{10}

\bibitem{ohloh}
Jason Allen, Scott Collison, Robin Luckey, and Andy Verprauskus.
\newblock {\em Ohloh}.
\newblock ohloh.net, 2008.
\newblock \url{http://www.ohloh.net/}, viewed in January 2008.

\bibitem{tomcat}
{Apache Foundation}.
\newblock {Apache Jakarta Tomcat}.
\newblock [online], 1999--2010.
\newblock \url{http://jakarta.apache.org/tomcat/index.html}.

\bibitem{aspectj}
{AspectJ Contributors}.
\newblock {\em AspectJ: Crosscutting Objects for Better Modularity}.
\newblock eclipse.org, 2007.
\newblock \url{http://www.eclipse.org/aspectj/}.

\bibitem{maya}
{Autodesk}.
\newblock {Maya}.
\newblock [digital], 2008--2010.
\newblock \url{autodesk.com}.

\bibitem{blender}
{Blender Foundation}.
\newblock {Blender}.
\newblock [online], 2008--2010.
\newblock \url{http://www.blender.org}.

\bibitem{mllessequal}
Fancois Bourdoncle and Stephan Merz.
\newblock On the integration of the functional programming, class-based
  object-oriented programming, and multi-methods.
\newblock Technical report, Centre de Mathemathiques Appliquees, Ecole des
  Mines de Paris and Institut fur Informatik, Technische Universitat Munchen,
  October 1996.

\bibitem{comparative-pls-3rd}
Robert~G. Clark.
\newblock {\em Comparative Programming Languages}.
\newblock Addison-Wesley, 3 edition, November 2000.
\newblock {ISBN:} 978-0201710120.

\bibitem{java-reflection}
Dale Green.
\newblock Java reflection {API}.
\newblock {Sun Microsystems, Inc.}, 2001--2005.
\newblock \url{http://java.sun.com/docs/books/tutorial/reflect/index.html}.

\bibitem{ISWIM}
P.~J. Landin.
\newblock The next 700 programming languages.
\newblock {\em Communications of the {ACM}}, 9(3):157--166, 1966.

\bibitem{louden97}
Kenneth~C. Louden.
\newblock {\em Compiler Construction: Principles and Practice}.
\newblock PWS Publishing Company, 1997.
\newblock {ISBN} 0-564-93972-4.

\bibitem{sed}
Lee~E. McMahon, Paolo Bonzini, Aur'{e}lio~M. Jargas, Eric Pement, Tilmann
  Bitterberg, Yao-Jen Chang, Yiorgos Adamopoulos, et~al.
\newblock \texttt{sed} -- stream editor for filtering and transforming text.
\newblock ftp://ftp.gnu.org/pub/gnu/sed and http://sed.sf.net, 1973--2006.
\newblock \url{http://sed.sourceforge.net/}, last viewed May 2008.

\bibitem{fcpp1}
Brian McNamara and Yannis Smaragdakis.
\newblock Functional programming in {C++} using the {FC++} library.
\newblock {\em SIGPLAN Notices}, 36(4):25--30, 2001.

\bibitem{isabelle-hol-tutorial}
Tobias Nipkow, Lawrence~C. Paulson, and Markus Wenzel.
\newblock {\em Isabelle/HOL: A Proof Assistant for Higher-Order Logic}, volume
  2283.
\newblock Springer-Verlag, November 2007.
\newblock \url{http://www.in.tum.de/~nipkow/LNCS2283/}, last viewed: December
  2007.

\bibitem{paquet-comp6411-w10-lecture-notes}
Joey Paquet.
\newblock Course notes for {COMP6411}, winter 2010.
\newblock Department of Computer Science and Software Engineering, Concordia
  University, Montreal, Canada, 2010.
\newblock [online].

\bibitem{isabelle}
Lawrence~C. Paulson and Tobias Nipkow.
\newblock Isabelle: A generic proof assistant.
\newblock University of Cambridge and Technical University of Munich, 2010.
\newblock \url{http://isabelle.in.tum.de/}, last viewed: February 2010.

\bibitem{fcpp2}
Yannis Smaragdakis and Brian McNamara.
\newblock {FC++}: Functional tools for object-oriented tasks.
\newblock {\em Softw., Pract. Exper.}, 32(10):1015--1033, 2002.

\bibitem{servlets}
{Sun Microsystems, Inc.}
\newblock Java servlet technology.
\newblock [online], 1994--2005.
\newblock \url{http://java.sun.com/products/servlets}.

\bibitem{jsp}
{Sun Microsystems, Inc.}
\newblock {JavaServer} pages technology.
\newblock [online], 2001--2005.
\newblock \url{http://java.sun.com/products/jsp/}.

\bibitem{flex}
{Various Contributors} and {the Flex Project}.
\newblock \texttt{flex}: The fast lexical analyzer.
\newblock [online], 1987--2008.
\newblock \url{http://flex.sourceforge.net/}, viewed in January 2008.

\bibitem{bison}
{Various Contributors} and {the GNU Project}.
\newblock Bison -- {GNU} parser generator.

\bibitem{gcc}
{Various Contributors} and {the GNU Project}.
\newblock {GNU Compiler Collection (GCC)}.
\newblock [online], 1988--2009.
\newblock \url{http://gcc.gnu.org/onlinedocs/gcc/}.

\bibitem{aspects-mem-management-java-cpp}
Emil Vassev and Joey Paquet.
\newblock Aspects of memory management in {Java} and {C++}.
\newblock In Hamid~R. Arabnia and Hassan Reza, editors, {\em Software
  Engineering Research and Practice}, pages 952--958. CSREA Press, 2006.

\bibitem{javacc}
Sreenivasa Viswanadha and {Contributors}.
\newblock Java compiler compiler ({JavaCC}) - the {Java} parser generator.
\newblock [online], 2001--2008.
\newblock \url{https://javacc.dev.java.net/}.

\bibitem{wikipedia}
Jimmy Wales, Larry Sanger, and {other authors from all over the world}.
\newblock Wikipedia: The free encyclopedia.
\newblock [online], Wikimedia Foundation, Inc., 2001--2010.
\newblock \url{http://wikipedia.org}.

\end{thebibliography}

\printindex

\end{document}